\documentclass[12pt]{JHEP3}

\usepackage{amssymb}
\usepackage{amsmath}
\usepackage{graphicx}
\usepackage{latexsym}

\newcommand{\psla}{p\kern-.45em/}

\newcommand{\be}{\begin{equation}}
\newcommand{\ee}{\end{equation}}
\newcommand{\bea}{ \begin{eqnarray} }
\newcommand{\eea}{ \end{eqnarray} }
\newcommand{\beqstar}{ \begin{eqnarray*} }
\newcommand{\eeqstar}{ \end{eqnarray*} }

\newcommand{\sla}[1]{\not\!#1}

\title{\Large Two jets and missing $E_T$ signature to determine the spins of new particles}

\author{Mihoko M. Nojiri\\
Theory Group, KEK,1-1 Oho, Tsukuba, Ibaraki 305-0801, Japan\\
The Graduate University for Advanced Studies (Sokendai), \\
1-1 Oho, Tsukuba, Ibaraki 305-0801, Japan\\
Institute for the Physics and Mathematics of the Universe, \\
The University of Tokyo, Chiba, 277-8583, Japan\\
}
\author{Jing Shu\\
Institute for the Physics and Mathematics of the Universe, \\
The University of Tokyo, Chiba, 277-8583, Japan}

\preprint{
KEK-TH-1435\\
IPMU10-0233
 }

\abstract{
We consider the spin determination of new colored particles in the missing energy plus jets channel at the early stage of the discovery. We use a three site moose model to describe the low energy Lagrangian of all same spin partner (LHT or UED like) models and check the gauge invariance of the amplitude. For the benchmark production and decay channel $pp \rightarrow U^{(R)} U^{(R)} \rightarrow u u B_H B_H$, in contrast to those in supersymmetric models, there are spin correlations which affect the polar and azimuthal angle distributions of the quarks from the heavy partner $U^{(R)}$ decay. We show such effects would be visible in the $E_{\rm T miss} / M_{\rm eff}$ distribution and the reconstructed azimuthal angle correlation using MAOS reconstruction.
}

\begin{document}
\section{Introduction}

The existence of dark matter in our Universe is solid evidence of new physics beyond the standard model. Dark matter must consist of a new elementary particle, whose lifetime is much longer than the age of our Universe. One way to incorporate the dark matter candidate beyond the Standard Model (SM) is to introduce a parity structure. This parity structure can be incorporated in well-studied models such as supersymmetry (SUSY) (R-parity), little Higgs  (LHT) (T-parity) \cite{Cheng:2004yc}, and universal extra dimensions  (UED) (Kaluza Klein parity) \cite{Appelquist:2000nn, Cheng:2002ab, Cheng:2002iz}. Under such a parity transformation, all the SM particles  have even parity, their partners have odd parity, and the interaction vertices multiplicatively conserve the parity. The lightest parity odd particle becomes stable, and 
becomes invisible in collider searches. 



Observation of large missing transverse momentum at a collider 
would raise further questions about the properties of the underlying physics. 
The signatures of SUSY and same spin-partner models are quite similar \cite{Hubisz:2008gg, Hallenbeck:2008hf}. In supersymmetric models, scalar quarks and scalar leptons have spin 0 while in LHT and UED, quark and lepton partners have spin $1/2$. The gauge boson partners are spin 1/2 in supersymmetric models, while they are heavy gauge bosons in LHT and UED. Therefore, it is critical to determine the spin of partners of the SM fields in order to distinguish different types of theories. To distinguish the spin structure of SM partners,  we need to study a quantity that can be measured in the early stage of the Large Hadron Collider (LHC). At a hadron collider, the main production channels contain colored objects, whose main decay channels involve jets.  The total cross section may serve as an initial hint to the spin of the odd partners \cite{Kane:2008kw}. However, at this stage of discovery, small statistics make it impossible to reconstruct the mass, which the cross section depends on strongly. 

The spin-dependence of decay patterns of new particles is useful in
determining the spin structure. The decay distributions depend on the spins of the
initial, intermediate and final state particles. For example, in the two leptons + jets + $E_{\rm Tmiss}$ channel at the LHC, the spin dependence appears in the jet invariant mass
distributions. The simple  $m_{ll}$ distribution of  $d \Gamma/m_{ll} \propto m_{ll}$
arising  from a $\tilde{q}\rightarrow \chi^0_2 j\rightarrow \tilde{l} jl\rightarrow \chi^0_1jll $ cascade decay in SUSY reflects the fact that the slepton is a scalar. In addition, the charge asymmetry in the $m_{jl}$ distribution arises from the  polarization \cite{Barr:2004ze} \cite{Goto:2004cpa} of the  spin 1/2 neutralinos. The polarization arises because the LHC is a proton-proton collider, so squarks are produced more often than anti squarks.

The difference between the invariant mass distributions of SUSY signatures and those of same spin partner models have been studied previously \cite{Smillie:2005ar, Datta:2005zs, Athanasiou:2006ef, Wang:2006hk, Burns:2008cp}. The spin and interaction dependence is sizable, but  such studies at the LHC may not be straightforward, because of the difficulty of  selecting the correct jets and leptons. For a relatively clean $jll$ signature, the decay branching ratio of $\tilde{q} \rightarrow \chi^0_2$ is typically only around 30\%,  and $\chi^0_2 \rightarrow \tilde{l} \rightarrow \chi^0_1$ is typically less than 10\%. It is not the dominant decay channel and is therefore statistically limited. To distinguish the spin structure of quark partners in the relatively early running of LHC, we need to focus on the polarization dependence of the
dominant decay channel.

The other  useful information lies  in the spin correlation  of particles from different decay chains. Here invariant mass distributions are not the useful quantity, rather the distributions of the polar angle and the azimuthal angle correlations \cite{Buckley:2007th, Buckley:2008pp, Boudjema:2009fz} of decays. Due to the uncertainty of initial parton energy, it is not easy to see the angular correlation exactly at the LHC, but the effect should exist in various distributions in cases with  strong angular correlations.



In this paper, we study the spin correlations of the $q q \rightarrow Q Q$ process where $Q$ is a spin 1/2 quark partner. We use a three-site deconstructed model which universally parametrizes low-energy effects in popular same spin partner (LHT or UED like) models with gauge boson and fermion partners. With such a simplified gauge symmetry breaking sector, it is straightforward to  check the gauge invariance of the amplitude which guarantees good high energy behavior. In our model, the pair production and the decay of the spin 1/2 quark partner can have a  spin correlation which is in contrast to the isotropic decays of scalar quarks. Among the possible signatures, the two high $p_T$ jet + missing $E_T$ +$ X$ channel is  promising, where $ X$ is an arbitrary number of jets and leptons. The quark partners can directly decay into a light stable gauge boson partner and a quark, and the two high $p_T$ jets come from their direct decays. It is essential that the jets coming from the two body decay tend to have a large $p_T$ because it makes their  selection much easier. We stress that the channel we are proposing, which is one of the main discovery channels at the LHC, could be the most promising one to see  spin correlations. Once the integrated luminosity becomes closer to 1 fb$^{-1}$, the LHC will be able to access partners with a production cross section above O(1) pb$^{-1}$. Therefore we 
propose simple distributions that are sensitive to the spin correlations, so that 
we would be able to determine the spin structure as early as possible. 



The paper is organized as follows. In Section \ref{sec:model}, we introduce a three site model  which gives us the  low energy spectra of the massive gauge boson and fermion partners without model parameter constraints. We double check the gauge invariance of the amplitudes which guarantees their good high energy behavior. In Section \ref{sec:partondis}, we use two sample spectra in parton level simulations and present the polar angle and azimuthal angle dependences. We also calculate the amplitudes in different helicity channels. In Section \ref{sec:jetlevel}, we demonstrate how the jet level distributions are affected by the spin correlations. Section \ref{sec:conclusion} contains our conclusions.

\section{Quark partner production in a three-site moose model.}
\label{sec:model}

There are various models which contain SM ``partners" with the same spin and an odd $Z_2$ parity. Examples such as 
the universal extra dimension model ( ``bosonic supersymmetry" ) \cite{Appelquist:2000nn, Cheng:2002ab, Cheng:2002iz}, its variants with split mass spectra \cite{Park:2009cs, Flacke:2008ne}, little Higgs models with T-parities \cite{Cheng:2004yc} or even the warped models with KK parities belong to this group \cite{Agashe:2007jb}. In the process $pp \rightarrow u_1 u_1$ followed by $u_1$  decay into a quark and light stable particle (+ softer jets or leptons), one can safely neglect the effects of electroweak symmetry breaking, and all the massive particles gain their masses through the extra gauge symmetry breaking or the compactification with bulk gauge symmetry. From the dimensional deconstruction \cite{ArkaniHamed:2001ca, Hill:2000mu} point of view, all of these models can be  described by the three site moose diagram. Another reason we choose the three site moose digram here is because we need to describe a Goldstone boson with relatively simple mass mixings, which are needed in the $R_\xi$ gauge to check the gauge invariance of the amplitude and the high energy behavior. On the other hand, it would be very cumbersome if we were to use the the fifth component of the gauge boson in the language of extra dimensions.

\begin{figure}
  \includegraphics[width=8cm]{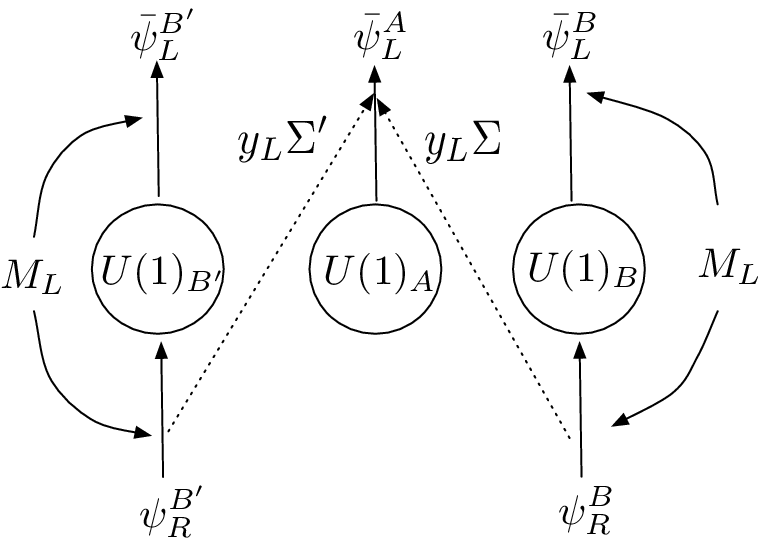}
    \includegraphics[width=8cm]{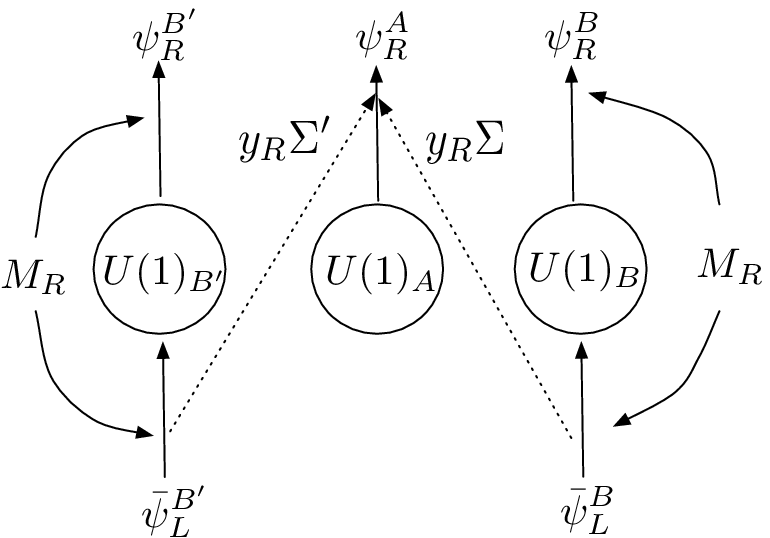}
  \caption{The moose diagrams for the three site split-UED model with a single 5D Dirac Fermion.
An arrow into a site means that the particle transforms under the fundamental representation of the relevant site and an arrow out of a  site means that the particle transforms under the anti-fundamental representation. The solid lines stand for Weyl fermions and the dashed lines represent scalars. The left  panel shows how to get a left-handed chiral zero mode and the lower right shows how to get a right-handed chiral zero mode.}
  \label{fig:3site}
\end{figure}

We consider a three site $U(1)$ moose model which mimics the low energy theory for the scattering $uu \rightarrow u_1 u_1$ in any kind of same spin partner (LHT or UED like) model. For the non-abelian case, our result would only differ by a color factor. The moose diagram is presented in Fig \ref{fig:3site}.  The gauge couplings in each moose are $g_{B^\prime}$, $g_A$ and $g_B$, and we set $g_B = g_{B^\prime}$ so the model has a $Z_2$ parity. The fermions charged under the gauge groups $U(1)_B$ and $U(1)_{B^\prime}$ have equal Dirac mass terms  $-M \bar{\psi}_L^B \psi_R^B + \rm{h. c.}$ and $-M \bar{\psi}_L^{B^\prime} \psi_R^{B^\prime} + \rm{h. c.}$. The two bifundamental scalar fields $\Sigma$ and $\Sigma^{\prime}$, which we will call the ``link" fields, are charged under gauge groups $U(1)_{B^\prime}$, $U(1)_{A}$ and $U(1)_{A}$, $U(1)_{B}$ with charge $(1,-1)$ and $(-1, 1)$ respectively. They couple to the fermions through the Yukawa couplings with the same coupling strength $y \bar{\psi^A_L} \Sigma \psi^B_R + \rm{h.c.}$ and $y \bar{\psi^A_L} \Sigma \psi^{B^\prime}_R + \rm{h.c.}$.


The link fields get their vacuum expectation values $ \langle \Sigma_{i\bar{k}} \rangle =  \langle \Sigma^{\prime}_{i\bar{k}} \rangle = u \delta_{i\bar{k}}$ and spontaneously break $U(1)_{B^\prime}$, $U(1)_A$ and $U(1)_B$ into the diagonal group $U(1)_0$. The kinetic term for the link fields $\rm{Tr}[(D_{\mu} \Sigma)^{\dag} (D_{\mu} \Sigma)]$ + $\rm{Tr}[(D_{\mu} \Sigma^\prime)^{\dag} (D_{\mu} \Sigma^\prime)]$ generates the mass terms for the massive gauge bosons. 
The mass matrix of the gauge bosons is  
\begin{eqnarray}
\begin{pmatrix}
B_{\mu} & A_{\mu} & B^\prime_{\mu}
\end{pmatrix}
\begin{pmatrix}
g_B^2 u^2 & - g_A g_B u^2 & 0 \\
- g_A g_B u^2 & 2 g_A^2 u^2 & -g_A g_B u^2 \\
0 & -g_A g_B u^2 & g_B^2 u^2
\end{pmatrix} 
\begin{pmatrix}
B_{\mu} \\ A_{\mu} \\ B^\prime_{\mu} 
\end{pmatrix}
 \label{G_MG} \ .
\end{eqnarray}


We can find the mass eigenstates of the gauge bosons in terms of the gauge eigenstates 
\begin{eqnarray}
\begin{pmatrix}
G^0_{\mu} \\ G^1_{\mu} \\ G^2_{\mu}
\end{pmatrix}
=
\begin{pmatrix}
s_g / \sqrt{2} & c_g & s_g / \sqrt{2} \\
- 1 / \sqrt{2} & 0 & 1 / \sqrt{2} \\
c_g / \sqrt{2}  & -s_g & c_g / \sqrt{2} 
\end{pmatrix} 
\begin{pmatrix}
B_{\mu} \\ A_{\mu} \\ B^\prime_{\mu}
\end{pmatrix}
\ .
 \label{G_TMG}
\end{eqnarray}
The inverse transformation between the gauge eigenstates and mass eigenstates is
\begin{eqnarray}
\begin{pmatrix}
B_{\mu} \\ A_{\mu} \\ B^\prime_{\mu}
\end{pmatrix}
=
\begin{pmatrix}
s_g / \sqrt{2} & - 1 / \sqrt{2} & c_g / \sqrt{2} \\
c_g  & 0 & -s_g  \\
s_g / \sqrt{2}  & 1 / \sqrt{2} & c_g / \sqrt{2} 
\end{pmatrix} 
\begin{pmatrix}
G^0_{\mu} \\ G^1_{\mu} \\ G^2_{\mu}
\end{pmatrix}
\ ,
 \label{G_TGM}
\end{eqnarray}
where $s_g \equiv { \sqrt{2} g_A} / g$, $c_g \equiv g_B / g$ and $g \equiv \sqrt{2 g_A^2 + g_B^2}$. The corresponding gauge boson masses for $G^0_\mu$, $G^1_\mu$, $G^2_\mu$ are 0, $g_B u$ and $g u$. 

Similarly, the gauge invariant Dirac mass term and the Yukawa interactions give the fermion masses. The fermion mass matrix is 
\begin{eqnarray}
\begin{pmatrix}
\psi_L^B & \psi_L^A & \psi_L^{B^\prime}
\end{pmatrix}
\begin{pmatrix}
M_L & y_L u  & 0 \\
0 & 0 & 0 \\
0 & y_L u & M_L
\end{pmatrix} 
\begin{pmatrix}
\psi_R^B \\ \psi_R^A \\ \psi_R^{B^\prime}
\end{pmatrix}
 \label{F_MG}
\end{eqnarray}
We can also find the mass eigenstates of the fermions in terms of the gauge eigenstates 
\begin{eqnarray}
\begin{pmatrix}
\psi^0_{L} \\ \psi^1_{L} \\ \psi^2_{L}
\end{pmatrix}
=
\begin{pmatrix}
s_f / \sqrt{2} & - c_f & s_f / \sqrt{2} \\
- 1 / \sqrt{2} & 0 & 1 / \sqrt{2} \\
c_f / \sqrt{2}  & s_f & c_f / \sqrt{2} 
\end{pmatrix} 
\begin{pmatrix}
\psi_L^B \\ \psi_L^A \\ \psi_L^{B^\prime}
\end{pmatrix}
\ ,
 \label{F_LTMG}
\end{eqnarray}
\begin{eqnarray}
\begin{pmatrix}
\psi^0_{R} \\ \psi^1_{R} \\ \psi^2_{R}
\end{pmatrix}
=
\begin{pmatrix}
0  & 1 & 0 \\
- 1/ \sqrt{2}  & 0 & 1 / \sqrt{2} \\
1/ \sqrt{2}  & 0 & 1 / \sqrt{2} 
\end{pmatrix} 
\begin{pmatrix}
\psi_R^B \\ \psi_R^A \\ \psi_R^{B^\prime}
\end{pmatrix}
\ .
 \label{F_RTMG}
\end{eqnarray}
The inverse transformation between the gauge eigenstates and mass eigenstates is
\begin{eqnarray}
\begin{pmatrix}
\psi_L^B \\ \psi_L^A \\ \psi_L^{B^\prime}
\end{pmatrix}
=
\begin{pmatrix}
s_f / \sqrt{2} & - 1 / \sqrt{2}  & c_f / \sqrt{2} \\
- c_f & 0 & s_f  \\
s_f / \sqrt{2}  & 1 / \sqrt{2} & c_f / \sqrt{2} 
\end{pmatrix} 
\begin{pmatrix}
\psi^0_{L} \\ \psi^1_{L} \\ \psi^2_{L}
\end{pmatrix}
\ ,
 \label{F_LTGM}
\end{eqnarray}
\begin{eqnarray}
\begin{pmatrix}
\psi_R^B \\ \psi_R^A \\ \psi_R^{B^\prime}
\end{pmatrix}
=
\begin{pmatrix}
0  & - 1/ \sqrt{2}  & 1/ \sqrt{2} \\
1 & 0 & 0 \\
0  & 1 / \sqrt{2} & 1 / \sqrt{2} 
\end{pmatrix} 
\begin{pmatrix}
\psi^0_{R} \\ \psi^1_{R} \\ \psi^2_{R}
\end{pmatrix}
\ .
 \label{F_RTGM}
\end{eqnarray}
Here, the fermion mixing angle is defined as $s_f \equiv { \sqrt{2} y u} / \sqrt{M^2 + 2 (y u)^2}$ and $c_f \equiv $$ {M} / $ $\sqrt{M^2 + 2 (y u)^2}$, and  the corresponding fermion masses for $\psi^0$, $\psi^1$, $\psi^2$ are 0, $M$, $\sqrt{M^2 + 2 (y u)^2}$. 

In this case, the model contains  one gauge boson sector and one chiral fermion sector. The gauge boson sector contains one parity even massless state (0 mode), one parity odd massive state (1st mode) and one parity even massive state (2nd mode); The chiral fermion sector  contains one parity even left-handed chiral fermion (0 mode), one lighter parity odd massive state (1st mode) and one lighter parity odd massive state (2nd mode). In order to mimic the SM and cancel the anomaly, one can also introduce another fermion sector which has a parity even right-handed chiral fermion as in the right panel of Fig. \ref{fig:3site}.

For the Goldstones, the interactions with  fermions are
\bea
\mathcal{L} &=& y \bar{\psi_L^A} \Sigma \psi_R^B + y \bar{\psi_L^A} \Sigma^\prime \psi_R^{B^\prime} + h. c. \nonumber \\
& \supset &  \frac{c_f y}{\sqrt{2}} \bar{\psi_L^0} \Sigma \psi_R^1 - \frac{c_f y}{\sqrt{2}} \bar{\psi_L^0} \Sigma^\prime \psi_R^1 + h. c. \nonumber \\
& \supset & i \frac{c_f y}{\sqrt{2}}   \bar{\psi_L^0} \psi_R^1 \pi - i \frac{c_f y}{\sqrt{2}}  \bar{\psi_R^1} \psi_L^0 \pi - i \frac{c_f y}{\sqrt{2}} \bar{\psi_L^0} \psi_R^1 \pi^\prime + i \frac{c_f y}{\sqrt{2}}  \bar{\psi_R^1} \psi_L^0 \pi^\prime \nonumber \\
 & \supset & - i {c_f y}  \bar{\psi_L^0} \psi_R^1 \pi_1 + i {c_f y} \bar{\psi_R^1} \psi_L^0 \pi_1  
  \ , 
\eea
where the symmetry breaking of $\sigma$ and $\sigma'$ is linear
\bea 
\Sigma = (u+ \sigma) + i \pi \ . \nonumber \\
\Sigma^\prime = (u+ \sigma^\prime) + i \pi^\prime
\eea
and $G^1_\mu$ eats the linear combination $\pi_1 \equiv (\pi^\prime - \pi)/\sqrt{2}$. 

\begin{figure}[htbp]
 \begin{center}
\includegraphics[width=4cm]{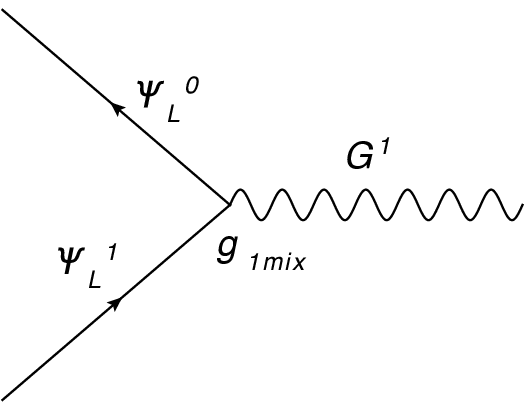}
\includegraphics[width=4cm]{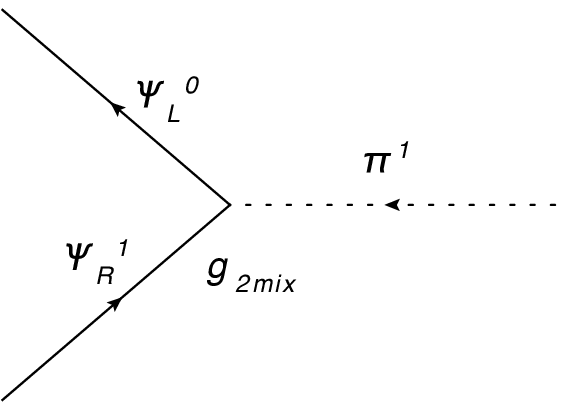}
     \caption{Feynman rules for the interactions used in calculating $ u u \rightarrow u_1 u_1$. The mixed coupling strength for the gauge boson and fermions is $g_{1 L} = g c_g s_g / \sqrt{2}$ and the mixed coupling strength for the Goldstone and fermions is $g^{\prime}_{1 L} = - y c_f $. }
 \label{fig:FeymRule3}
 \end{center}
\end{figure}

Now we can derive the Feyman rules for the model and calculate the full amplitude. The relevant Feynman rules that we use in the calculation are presented in Fig \ref{fig:FeymRule3}. The $\psi_L^1 - \psi_L^0 - G^1$ coupling  is $g_{1 L} = g c_g s_f / \sqrt{2}$, and the Goldstone $ \psi^1_R - \psi^0_L - \pi_1$  coupling is $g^{\prime}_{1 L} = -c_f y$. For the scattering process $u u \rightarrow u_1 u_1$, one must make sure that the calculated amplitude does not have bad high energy behavior. Indeed, as  t'Hooft and Veltman in their proof of the renormalizability of the electroweak theory \cite{'tHooft:1972fi}, it is the gauge invariance of the theory which guarantees the cancellation of bad high energy behavior of the amplitude \cite{Cornwall:1974km}. In order to check the gauge invariance of the amplitude, we adopt the general $R_\xi$ gauge and add the full amplitude including both massive gauge boson and Goldstone boson exchange. 


The amplitude coming from the t-channel exchange of the massive gauge boson in the $R_\xi$ gauge is 
\bea
i \mathcal{M} &=& \left (- i g \frac{s_f c_f } {\sqrt{2}} \right )^2 \bar{u} (p') \gamma_\mu P_L u(p) \frac{- i}{q^2 - m_G^2} \times (-)
\nonumber \\
& & \left( g^{\mu \nu} - \frac{q^\mu q^\nu}{q^2 - \xi m_G^2}(1- \xi) \right )   \bar{u} (k') \gamma_\nu P_L u(k) \ .
\eea
We can extract the gauge independent piece from the gauge boson propagator as 
\bea
\frac{-i}{q^2 - m_G^2} \left( g^{\mu \nu} - \frac{q^\mu q^\nu}{m_G^2} \right ) + \frac{- i}{q^2 - m_G^2 \xi} \left( \frac{q^\mu q^\nu}{m_G^2} \right ) \ .
\eea
So the additional gauge dependent piece can be written as 
\bea
i \mathcal{M_\xi} &=& \frac{1} {2} (g_B s_f)^2 \bar{u} (p') P_L u(p) \frac{- i}{q^2 - m_G^2 \xi} \frac{- m_1 ^2}{ m_G^2} \bar{u} (k') P_L u(k)
\nonumber \\
& = & (y c_f)^2  \bar{u} (p') P_L u(p) \frac{ i}{q^2 - m_G^2 \xi} \bar{u} (k') P_L u(k) \ .
\eea
Notice that $m_1 = M = \sqrt{2} y u / t_f $ and $m_{G_1} = g_B u$.

The amplitude for the Goldstone interaction is 
\bea
i \mathcal{M_\xi} &=& (- c_f y)^2 \bar{u} (p') P_L u(p) \frac{ i}{q^2 - m_G^2 \xi}\times (-) 
\nonumber \\
& &  \bar{u} (k') P_L u(k) \ ,
\eea
which cancels the contribution from the gauge dependent piece from the t-channel gauge boson exchange. 

\section{Simplified Lagrangian and parton level distributions}
\label{sec:partondis}

With the above three site model, we can adjust the input parameters $M_L$, $g_A$, $g_B$, $y_L u$, $y_R u$, $M_R$ to get the free mass spectra for the massive gauge bosons and left-handed or right-handed chiral fermions and their massive vector partners. The relevant particles in same spin partner (LHT or UED like) models with the two jets plus missing energy signature are 
\begin{description}
\item{(1)}  a massive gluon partner $G_H$ or electroweak boson partner $W_H$.
\item{(2)} a light stable gauge boson $B_H$.
\item{(3)} light fermion partners $Q^{(L)}$, $Q^{(R)}$, $L^{(L)}$, $L^{(R)}$. (Here the superscript indicates the whether the fermion partner is partner of left-handed or the right-handed SM fermions, not the chirality of the partner.). 

\end{description}
All the fermion partner masses, gauge boson partner masses, and their interactions are free, but we assume the lightest partner is $B_H$, the partner of the SM $U(1)_Y$ gauge boson, and all partners have parity $-1$, so that  $B_H$ is stable. The other massive gauge boson partners, on the other hand, could be heavier or lighter than the massive fermion partners,  depending
on which  model parameters we choose. The interactions involving gauge boson and fermion partners are 
\begin{eqnarray}
{ L}_{int}&=&\int dx^4 \left[ g_{3 L} G^{\mu a}_H \bar{Q}^{(L)}_L T^a \gamma_{\mu} q_L+ g_{3 R} G^{\mu a}_H\bar{Q}^{(R)}_R T^a \gamma_{\mu} q_R + g_{2L} W^{\mu a}_H \bar{Q}^{(L)}_L T^a \gamma_{\mu}P_L q_L \right. \cr 
&& \left. + g_{1L} B^{\mu}_H Y_L \bar{Q}^{(L)}_L \gamma_{\mu} q_L+ g_{1R} B^{\mu}_H Y_R \bar{Q}^{(R)}_R \gamma_{\mu} q_R +({\rm Lepton \ part }) +h.c \right] 
\label{lag}
\end{eqnarray}
Here $g_{3 L/R}$, $g_{2 L}$, and  $g_{1 L/R}$ stand for the couplings between
the $SU(3)_c$, $SU(2)_L$, $U(1)_Y$ gauge boson partners and the SM left-handed
or right-handed fermions and their partners. For simplicity, we take them to be the same as those of the SM gauge bosons in the following. We do not write the Goldstone interactions which ensure the gauge invariance of the amplitude.

Since the parton distribution functions of the first generation quarks are much harder than those of anti-quarks or gluons for protons, the production cross section of quark partners is much larger than that of antiquark partners. Gluon partners can decay 
into quark partners or anti-quark partners, however their production cross section is still subdominant. 
Therefore, light quark partner production $pp\rightarrow 
Q^{(L)} Q^{(L)} / Q^{(R)} Q^{(R)}$ is the dominant production channel at the LHC. Notice that the subdominant production of $G_H G_H$, $G_H Q^{(R)} $ and $Q\bar{Q}$ may involve second level massive gauge boson exchanges in the s or t-channel unlike $QQ$ production. The cross section for those subdominant process may be significantly different from that of our simplified model.

The production and decay distributions of fermionic partners are quite different from those of scalar quarks in supersymmetric models due to spin correlations. Here we demonstrate how those spin correlations appear in different helicity amplitudes. The amplitude for $u^a(p_i,h_i)u^{a'}(p'_i,h'_i)\rightarrow U^{(R)b}(p_f,h_f)U^{(R)b'}(p'_f,h'_f)$ would be written as follows 
\begin{eqnarray}
i{\cal M}(uu\rightarrow U^{(R)}U^{(R)}) &=&i g'^2Y^2_u\delta_{ab}\delta_{a'b'}\frac{ \left(-g^{\mu\nu} +\frac{q^{\mu}q^{\nu}}{m_{B_H}^2} \right) }{q^2-m^2_{B_H}} \times \cr 
&&\times \bar{u}_{h_f}(p_f) \gamma_{\mu}P_R u_{h_i}(p_i) 
\bar{u}_{h'_f}(p'_f) \gamma_{\nu}P_R  u_{h'_i}(p'_i) \cr
&& +{\rm gluon \ partner 
\ exchange\  contributions}\cr
&&  + {\rm cross\  diagram }
\end{eqnarray}
where $p$'s are momenta, $h$'s are helicities of the particles, and $a, b$ are color indices. For the massive gauge boson propagator, we only write down the gauge independent piece since the rest of the pieces will be cancelled by the Goldstone boson exchange. We only show the $t$-channel $B_H$ exchange contribution and the $g_{\mu\nu}$ and $q^{\mu}q^{\nu}$ terms in the $B_H$ propagator. The extension to the full  amplitude is straightforward and for the numerical calculation we use the full amplitude.  The amplitude arising from the 
$g_{\mu\nu}$ term in the center of mass (CM) frame is
\begin{equation}
i{\cal M}_{1}=i g^{\prime 2}Y^2_u \delta_{ab}\delta_{a'b'}
\delta_{h_i,\frac{1}{2}}\delta_{\lambda_f,0} \frac{E^2_{CM}}{q^2-m^2_{B_H}}
\left[
2\delta_{h_f,\frac{1}{2}}\ \beta_f -(-)^{h^\prime_f+\frac{1}{2}} (1-\beta_f) 
\right] \ ,
\end{equation} 
which is non-zero for $h_i=h'_i=1/2$ and $\lambda_f \equiv h_f - {h'_f}=0$. On the other hand, the $q^{\mu} q^{\nu}/m^2_B$ term after using the equation of motion becomes 
\begin{equation}
i{\cal M}_{2}=-i {g'}^2Y^2_u\delta_{ab} \ 
\delta_{a'b'}\frac{m^2_{U^{(R)}}}{(q^2-m_B^2) m_B^2}
\bar{u}_{h_f}(p_f) P_R\  u_{h_i}(p_i)\  \bar{u}_{h'_f}(p'_f) P_R\  u_{h'_i}(p'_i) \ ,  
\end{equation}
which is enhanced by a factor of  $m_{U^{(R)}}^2 / m_{B_H}^2$. Note that  this coupling is Yukawa-like, and it  therefore  flips the helicity of the fermions in the relativistic limit. The amplitude in the CM frame is
\begin{eqnarray}
i{\cal M}_{2}&=&-i{g'}^{2} Y^2_u\delta_{ab}\delta_{a'b'}\delta_{h_i,\frac{1}{2}} 
\delta_{\lambda_f, 0}
\delta_{\lambda_i,0}\,
\frac{m^2_{U^{(R)}}}{m^2_{B_H}(q^2-m_{B_H}^2)}
\left(\frac{E_{CM}}{2}\right)^2 \cr
&& 
\ \ \ \times 
\left[\left(2 \delta_{{h'}_{f},-{\frac{1}{2}}} \, \beta_f + (-)^{\bar{h}_f+\frac{1}{2}} (1-\beta_f)  \right)
\right.
\cr 
&& \left. 
-\frac{3}{4}   d^1_{\lambda_i\lambda_f} (\cos\theta) e^{i(\lambda_i -\lambda_f)\phi}
\left(\delta_{\lambda_f , 0 } (1-\beta_f) + 2\sqrt{2}\vert\lambda_f\vert 
\frac{m_{Q^{(R)}}}{E_{CM}} + 
2\delta_{{h'}_f,-\frac{1}{2}} \delta_{\lambda_f, 0 }\,  \beta_f\right)  
\right]. \cr
&&
\end{eqnarray}
In the relativistic limit, $\beta_f\rightarrow 1$, the amplitude is dominated by 
$(h_f, h'_f)=(1/2,1/2)$ and $(-1/2, -1/2)$ contributions. When $m_{Q^{(R)}}/m_B\gg 1 $, the $h= -1/2$ amplitude is dominant due to the Yukawa nature of the coupling. 

The decay of polarized $Q^{(R)}$ ($Q^{(R)}\rightarrow q B_H$) is not isotropic. The amplitude is 
dominated by the $h=0$ component of $B_H$ if $m_{Q^{(R)}} \gg m_{B_H}$, therefore the decay distribution is essentially that of a spin $1/2$ particle decaying into a light spin $1/2$ particle and a scalar. Because of the chiral nature of the vertices such as $B_H \bar{U}_R \gamma_\mu P_R u_R$, the quark in the final state is right handed. For a quark partner with $h=1/2$, the decay amplitude is given by
\bea
i{\cal M_D} (U^{(R)} \rightarrow u B_H ) & \propto & \epsilon^{*\mu}\epsilon^{\nu}{\rm Tr} \left[\gamma_\mu P_R \not p_f  \gamma_{\nu}
P_R \frac{1+ \not n \gamma_5 }{2}(\not p_i+ m_{U^{(R)}}) 
\frac{1+\not n \gamma_5}{2}\right]  \nonumber \\
&=& \frac{2k_B\cdot p_f m_{U^{(R)}}}{m^2_B}(E_B-k_{B//}) , 
\eea
where $\not n$ is the polarization vector with $n^2=-1$  and $k_{//}= -n\cdot k_B$ 
and $E_B$ is the energy of $B_H$ in  the $Q^{(R)}$'s rest frame.    
Quarks from $Q^{(R)}$ decays with $h_Q=1/2$ tend to go in the direction of $Q^{(R)}$ helicity as a result of  helicity conservation. The distribution is proportional to $1+\cos\theta_{Qq}$ for $h_Q=1/2$ in the massless limit, where $\theta_{Qq}$ is 
the polar angle of the $u$ momentum in the $U^{(R)}$ rest frame, where the z axis coincides with the $U^{(R)}$ momentum in the CM frame. The decay distribution of $Q^{(L)}$ ($Q^{(L)} \rightarrow q B_H$ or $q W_H$) with $h=-1/2$ is the same as that of  $Q^{(R)}$, $d\Gamma/d\cos\theta_{Qq} \propto 1+\cos\theta_{Qq}$ in the massless limit. On the other hand, it is opposite for an antiparticle decay. Because the number of produced quark partners is greater than the number of anti-quark partners at the LHC, the net polarization effect still remains in the signature.  

We calculate the amplitude using Madgraph/MadEvent \cite{Alwall:2007st} assuming that the collision energy for the LHC is 7~TeV. The MAOS momenta parameter cards of the model are obtained by BRIDGE \cite{Meade:2007js}. Fully spin correlated 
event distributions are generated by calculating the amplitude 
at the level of the final state partons $qq\rightarrow QQ \rightarrow qqB_HB_H$. 
Event distributions without spin correlations 
are generated for comparison 
by interfacing the $qq\rightarrow QQ$ amplitude of Madgraph/Madevent 
to PYTHIA \cite{Sjostrand:2006za}. We interface the 
events into AcerDet \cite{RichterWas:2002ch} for detector simulation, but for jet reconstruction 
we adopt the Cambridge/Aahen algorithm using FASTJET \cite{Salam:2007xv}. The number of events is $O(10^4)$ before the cuts. For a typical production rate $\sim$ pb at 1 $fb^{-1}$ integrated luminosity, the real number of events in the data sample would be about ten times smaller. Nevertheless, the qualitative features that we show in the plots would be the same. 


In our simulation, we consider two sample points in the parameter space. We first choose the following spectra: The heavy gauge boson partner $G_H$ and fermion partner $Q^{(R)}$ have quasi-degenerate masses (we choose $Q^{(R)}$ to be  relatively lighter) and are much heavier than the lightest gauge boson partner $B_H$. In this case, one can easily identify the hard jets from a heavy $Q^{(R)}$ decaying into $B_H$. We consider two different values of the light $B_H$ mass, which has a large impact on the polarization of jets, as explained in the introduction, and illustrate how those different polarizations give us various different distributions.     
\begin{description}
\item{(A)}  $m_{U^{(R)}}=600$~GeV, $m_{G_H}=700$~GeV, $m_{B_H}=100$~GeV. In this case, the sum of the production cross section of  $pp\rightarrow Q^{(R)}  Q^{(R)}$, $Q^{(R)} 
\bar{Q}^{(R)}$ and $\bar{Q}^{(R)} \bar{Q}^{(R)}$   is 5.8~pb. The dominant production channel $Q^{(R)} Q^{(R)}$ has a cross section of  5~pb.
\item{(B)} $m_{U^{(R)}}=600$~GeV, $m_{G_H}=700$~GeV, but the lightest gauge boson mass is heavier, $m_{B_H}=200$~GeV. In this case, the production cross section is 2.7~pb, which shows that the $B_H$ exchange contribution is important at Point A. 
\end{description}
For the spectra we choose here, the contribution of $G_H$ exchange is larger than that of $B_H$ exchange at point B, while the $B_H$ and $G_H$ exchange contributions are of the same order at the point A. 

\begin{figure}
\begin{center}
\includegraphics[width=6.5cm]{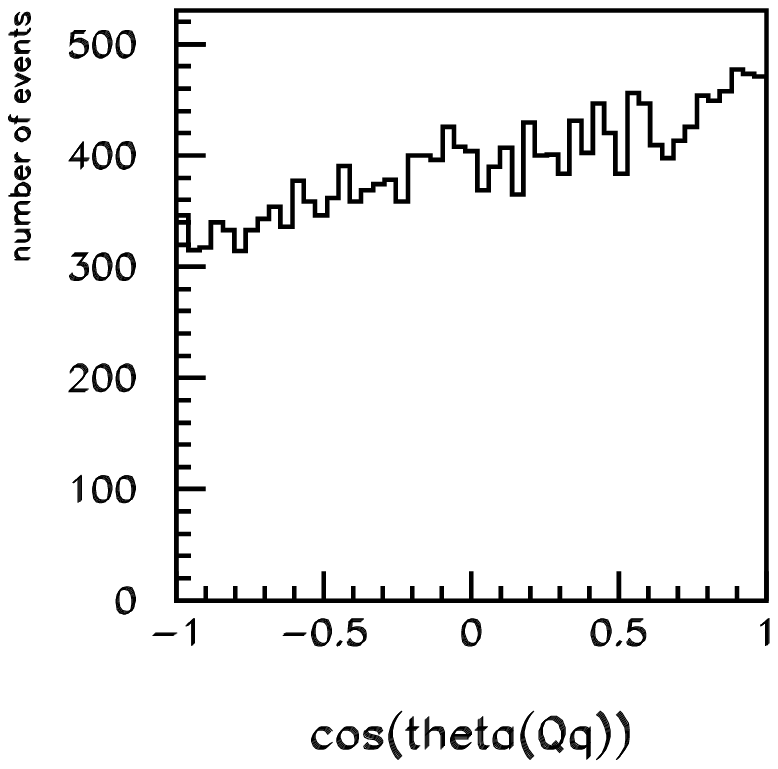}
\includegraphics[width=6.5cm]{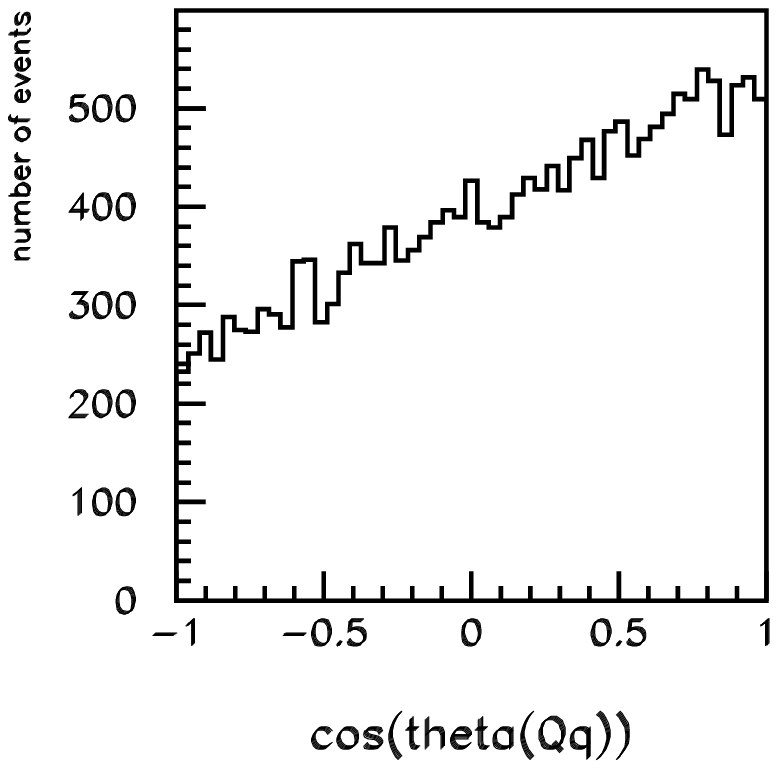} 
\end{center}
\caption{The $\cos\theta_{Qq}$ distributions at Point A with  $m_{B_H}=100$~GeV  (left) and at Point B with 
$m_{B_H}=200$~GeV (right). $m_{G_H}=700$~GeV and $m_Q=600$~GeV. All histograms correspond to $10^4$ events. }
\label{fig3}
\end{figure}

In Fig.~\ref{fig3}, we show  the $\cos\theta_{Qq}$ distribution for $pp\rightarrow U^{(R)}U^{(R)} \rightarrow uuB_H B_H$. We generate $10^4$ events using Madgraph, so each histogram contains $2\times 10^4$ entries. At Point A, the mass of $B_H$ in the $t$-channel exchange is small, therefore the Yukawa-like coupling coming from the $q^\mu q^\nu / m^2_{B_H}$ term in the $B_H$ propagator is enhanced, so there are equal order (1/2, 1/2) and (-1/2, -1/2) helicity states for $U^{(R)} U^{(R)}$, which tends to decrease the overall polar angle dependence. In the Point B with larger $B_H$ mass, the Yukawa coupling decreases. Only (1/2, 1/2) helicity states for $U^{(R)} U^{(R)}$ dominate and the overall polar angle dependence is stronger.

 
\begin{figure}[th] 
\begin{center}
\includegraphics[width=6.5cm]{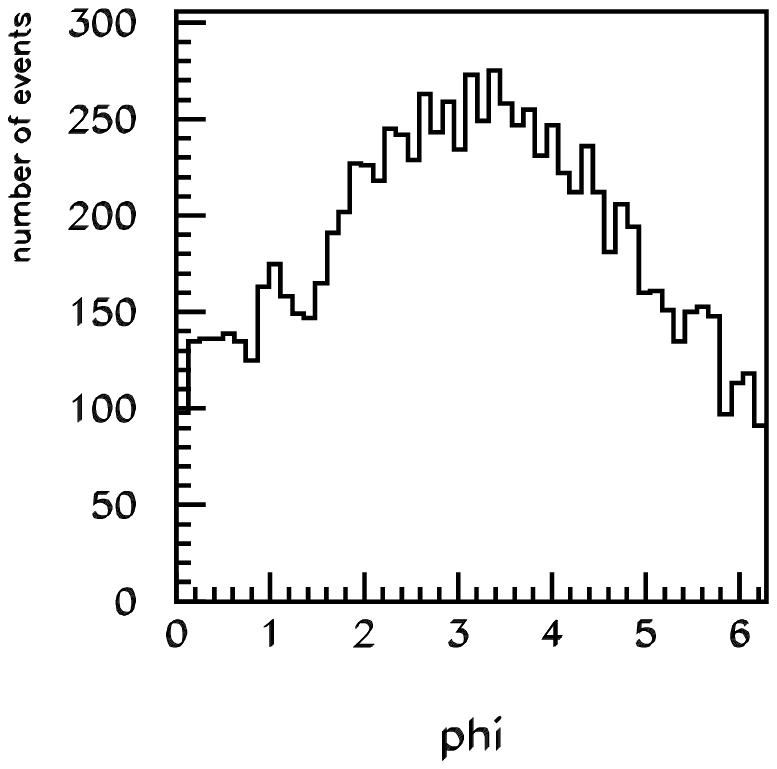}
\includegraphics[width=6.5cm]{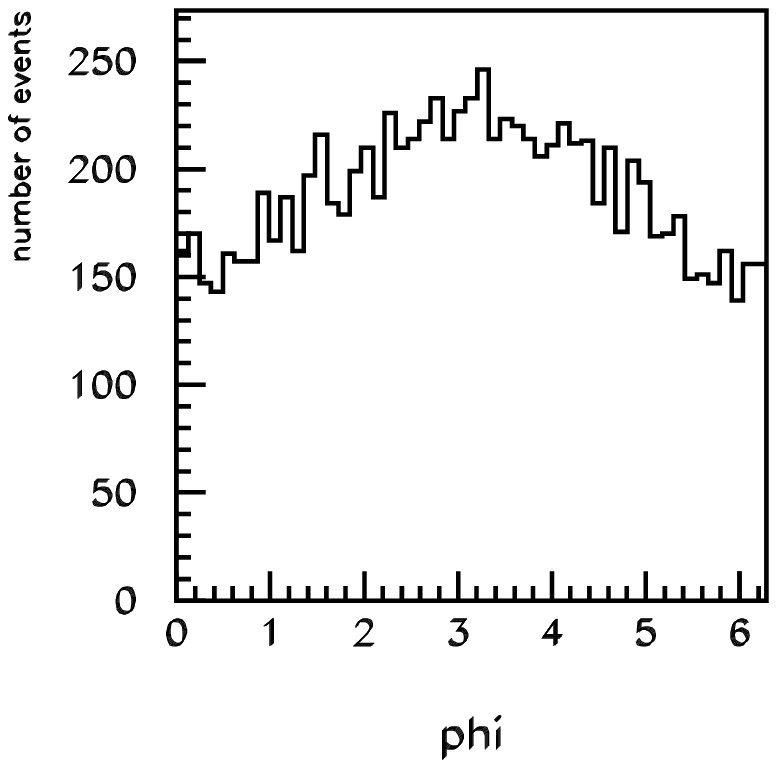}
\end{center}
\caption{The distributions of azimuthal angle difference $\phi$ of two $U^{(R)}$ decays in the events at Point A (left) and 
Point B (right). All histograms correspond to $10^4$ events.  
 } \label{fig4}
\end{figure}

We now discuss the azimuthal angle correlation in the $uu\rightarrow U^{(R)} U^{(R)} \rightarrow u u B_H B_H$ distribution \cite{Buckley:2007th, Buckley:2008pp}. The amplitude is 
expressed as the product of the production and decay amplitudes, 
\begin{equation}
iM(uu\rightarrow U^{(R)}U^{(R)}\rightarrow u u B_H B_H) 
= \sum_{h, h', \bar{h}, \bar{h}' }P_{h, h', \bar{h}, \bar{h}'}
(E_{\rm cm},\theta)  D_{h \bar{h}}(\theta^*, \phi^*)D_{h' \bar{h}'}(\theta'^*,\phi'^*) \ ,
\end{equation}
where $P_{h, h', \bar{h}, \bar{h}'} \equiv T_{h, h'} T^*_{\bar{h}, \bar{h}'}$ and $T_{h, h'}$ is the helicity amplitude of $uu\rightarrow U^{(R)}(h) U^{(R)}(h')$. The 
decay matrix  $D_{h, \bar{h}}(\theta^*, \phi^*)$
  for $U^{(R)}\rightarrow u B_H$ is a function of  $\theta^*$ and $\phi^*$. Here 
  $\theta^*$ is the polar angle of the momentum of $u$ in the rest frame of $U^{(R)}$, where the $z$ axis coincides with the momentum direction of $U^{(R)} $ in the CM frame (the production frame), and $\phi^*$ is the azimuthal angle relative to the production plane defined by the $u$ and $U^{(R)}$ momentum directions. The decay matrix has simple azimuthal-angle dependence,  
 \begin{equation}
 D_{h \bar{h}}(\theta^*, \phi^*)\equiv D_{h, \bar{h}}(\theta^*) e^{\mp i(h-\bar{h})\phi^*}.
 \end{equation}
 The factor of $\exp (\mp (h-\bar{h})\phi^*) $ is 1 if the production 
 amplitude is dominated by a single helicity state $h=1/2$ or $h=-1/2$, but if both $h=\pm 1/2$ amplitudes are sizable, the azimuthal angle correlation is important due to the large interference term. In that case the distribution has the form  $\propto 1+ a \cos(\phi)$ where $\phi$ is the difference of the two azimuthal angles. In Fig.~\ref{fig4}, we show the distributions of $\phi$ at Point A ($m_{B_H}=100$~GeV) and at Point B ($m_{B_H}=200$~GeV). The azimuthal angle correlation is stronger at Point A because the amplitudes for $(h, h')=(1/2,1/2)$ and $(h, h')=(-1/2,-1/2)$ are roughly equal at this point. The polar and azimuthal angle dependences are complementary to each other.  

\section{Detecting spin correlations in jet level distributions}
\label{sec:jetlevel}

In this section, we discuss the effects of spin correlations in jet level distributions.
Unlike at $e^+e^-$ colliders, at hadron colliders
it is not straightforward to see the effects by reconstructing the events exactly, because the collision energy of each event cannot be measured, 
and the lightest gauge boson partners are missed by the detector .  In this section, we point out a few distributions which are sensitive to the effects. 

\begin{figure} 
\begin{center}
\includegraphics[width=4.5cm]{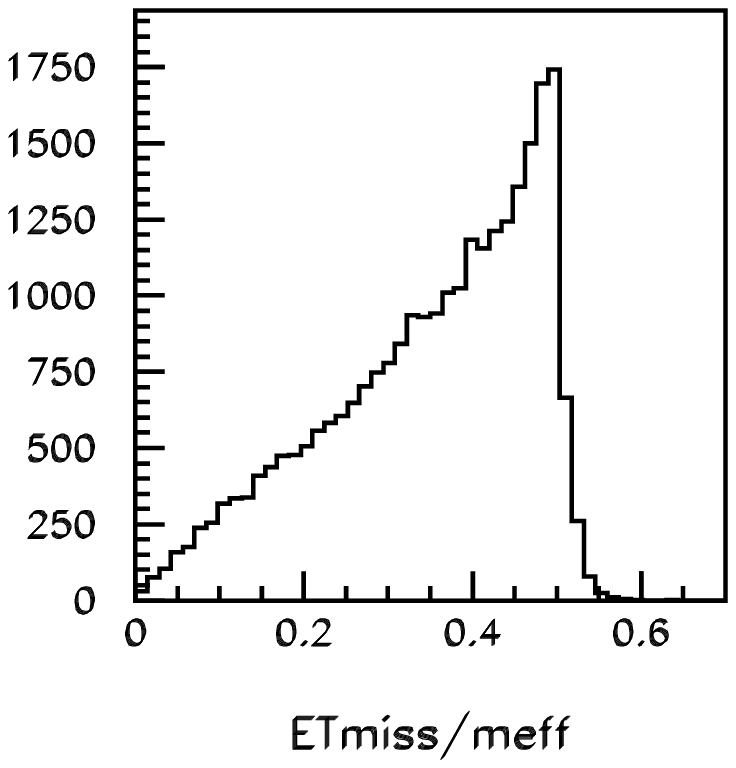}
\includegraphics[width=4.5cm]{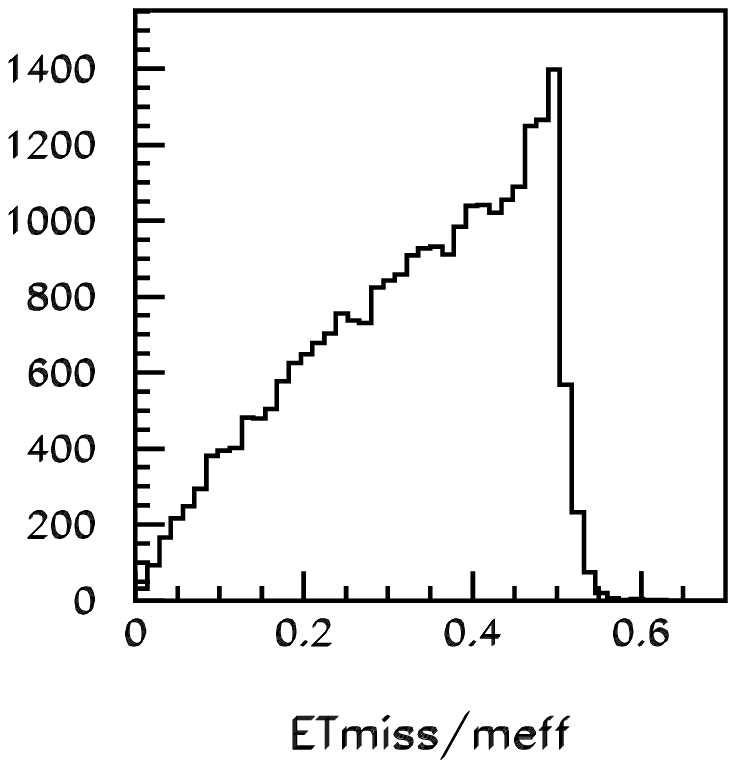}
\includegraphics[width=4.5cm]{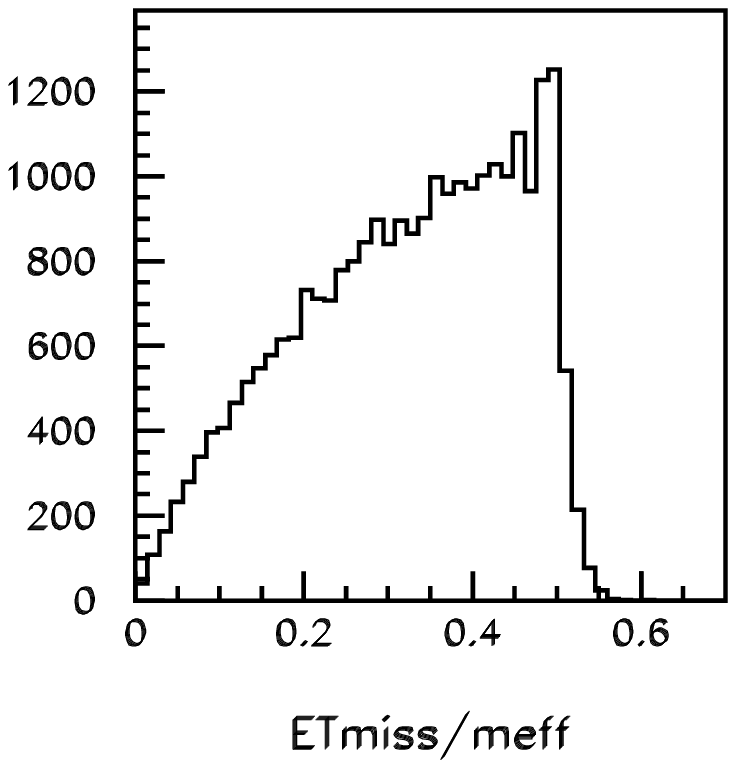}
\end{center}
\caption{Distributions of $E_{Tmiss}/M_{\rm eff}$ for $pp\rightarrow
 U^{(R)}U^{(R)}\rightarrow u_Ru_R B_H B_H$. Left; isotropic decays using PYTHIA at Point A. Middle; spin correlated decays at Point A. Right; spin correlated decay at Point B. All histograms correspond to $3\times 10^4$ events before the cuts. 
 } \label{fig5}
\end{figure}

In Fig.~\ref{fig5}, we show the distributions of $E_{\rm Tmiss}/M_{\rm eff}$ for 
 the  $pp \rightarrow U^{(R)} U^{(R)}$ process. 
  In the left figure we show the distribution at point A; here the $U^{(R)}$ decay is isotropic ($m_{B_H}=100$~GeV), which mimics SUSY squark production and decay. Fig.~\ref{fig5} (middle) is the fully spin correlated distribution for $pp \rightarrow uu B_H B_h$ at Point A,  and  Fig.~\ref{fig5} (right) is the distribution at Point B ($m_{B_H}=200$~GeV). Here  $M_{\rm eff}$ is defined as 
\begin{equation}
M_{\rm eff} \equiv E_{\rm Tmiss}+ \sum_i  p_{Ti} \ \ \ ,  
 \end{equation}
where the sum is over jets with $p_{T} > 50$ GeV.  $M_{\rm eff}$ is correlated with the typical CM energy of the collisions. 
The missing transverse energy $E_{T miss}$ is defined as 
\begin{equation}
E_{\rm Tmiss}=\sqrt{(P^{\rm miss }_X)^{2}+(P^{\rm miss}_Y)^{2}},
\end{equation}
where $-P^{\rm miss}_{X(Y)}$ is the sum of the transverse momenta of the reconstructed objects.  Up to the detector smearing and acceptance, it corresponds to the sum of the momenta of the invisible particles. 
 
\begin{figure} 
\begin{center}
\includegraphics[width=4cm]{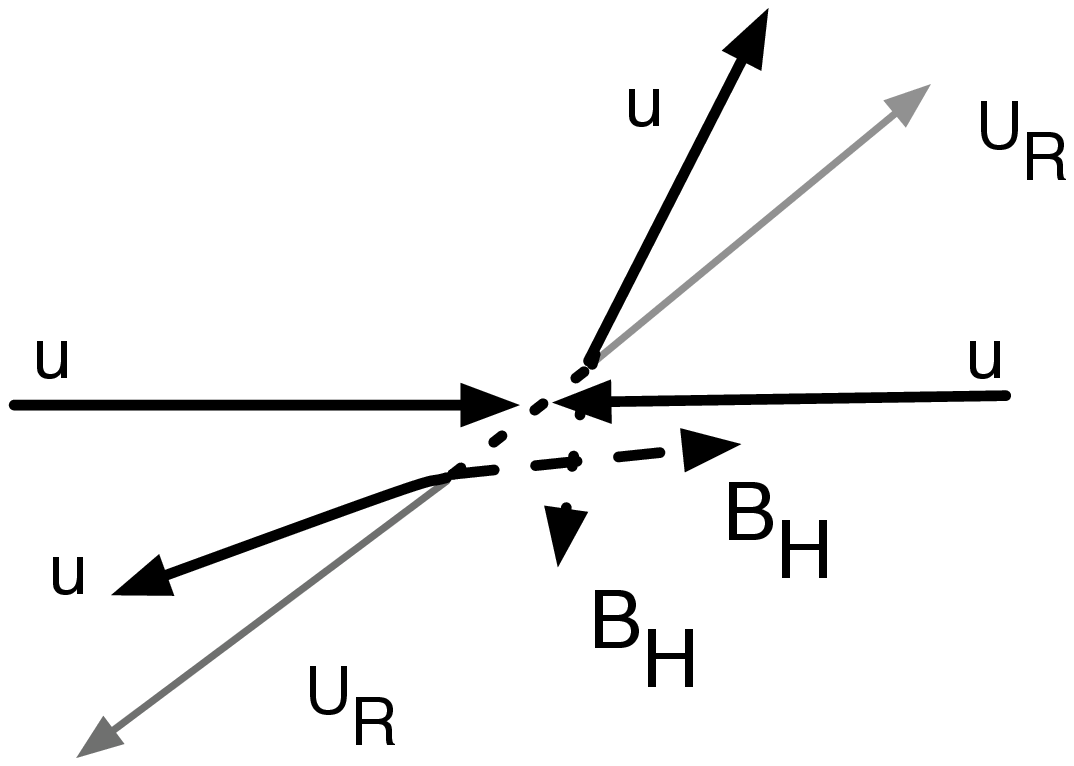}
\hskip 1cm
\includegraphics[width=4cm]{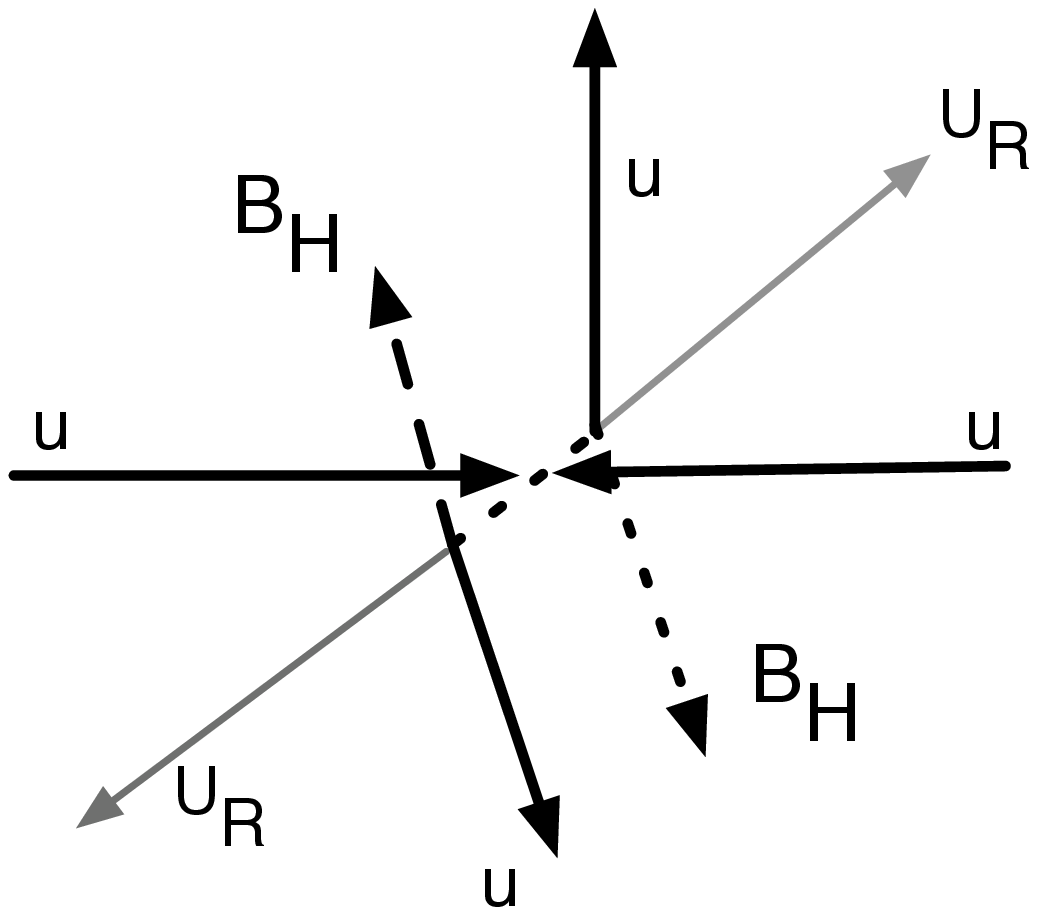}
\end{center}
\caption{Schematical description of event topologies of the process $U^{(R)}U^{(R)}\rightarrow u_Ru_R B_H B_H$. Left: A case of polarized quark partners (Point B). The $u$ quarks tend go in the same directions as their parent $U^{(R)}$'s go. Right: the case  with strong azimuthal angle correlation(Point A). Two $u$ quarks tend to go in back to back directions in the production frame. 
 } \label{fig6}
\end{figure}

The distribution of the events without spin correlation, (which is obtained by interfacing the $pp\rightarrow UU$ events to PYTHIA for isotropic decays) in Fig. \ref{fig5} (left) is clearly different from the spin  correlated decay. The spin correlated events tend to have smaller missing energy for fixed $M_{\rm eff}$. The typical decay topologies at Points A and B are shown in Fig.~\ref{fig6} (right) and Fig.~\ref{fig6} (left), respectively. The left figure shows a decay topology at Point B where the average polarization of the $U^{(R)}$ is right handed. 
The quarks tend to go in the directions of  the parent $U^{(R)}$ momenta, therefore the energy of 
$u$ ($B_H$) from  $U^{(R)}$ decay is boosted up (down) in average. The average 
$E_{\rm Tmiss}/M_{\rm eff}$, which is correlated with 
$P^T_{B_H}/m_{U^{(R)}}$, is smaller compared with that of isotropic 
decay. 

In Fig.~\ref{fig6} (right), we show the other decay pattern reducing the average 
$E_{\rm Tmiss}/M_{\rm eff}$, which is  represented by
the distribution at Point A. At this point,
the average polarization of $U^{(R)}$ is small, but the azimuthal angle correlation is 
stronger. In addition, the direction of $U^{(R)}$ production is peaked forwards because $m_{B_H}/m_{U^{(R)}}$ is small. Therefore the azimuthal angle distribution of $u$ in the lab frame and that in the production frame are similar. The transverse momenta of the $u$ quarks tend to be back to back in the  production frame, therefore the two $B_H$'s tend to be back to back both in the production plane and in the lab transverse plane. The $B_H$ transverse momenta cancel on average, so the total missing $p_T$ becomes smaller. 


Next we discuss reconstruction of the azimuthal angle correlations. The  
simplest  way 
to see this  for our process is to look into the $\phi$ distribution of the leading two jets in the Lab frame.  However, in this paper we use 
MAOS momentum
reconstruction \cite{Cho:2008tj, Cho:2009wh} to reconstruct the $\phi$ distribution. This is a method for estimating the directions of missing particle momenta using the $M_{T2}$ parameter \cite{Barr:2003rg}, which we explain below. 

The definition of $M_{T2}$ is expressed as follows \cite{Barr:2003rg}, 
\begin{equation}
M_{T2}=\min_{p^{T}_1+p^{T}_2=p_{\rm Tmiss} }\left[\max \left(
m_{T}(p^{(1)}_{\rm vis},p_1^{T}, m_{\rm test} ) ,m_{T}(p^{(2)}_{\rm vis},p^{T}_2, 
m_{\rm test})
\right) \right]. 
\end{equation}
Here $p^{(i)}_{\rm vis}$ are the momenta of the two visible objects
 in the events. We adopt the two highest 
 $p_T$ jets in each event, and $p_{1,2}$ are those of the two invisible objects as $p^{(i)}_{\rm vis}$, and denote the calculated $M_{T2}$ by 
  $m_{T2}$ or  $m_{T2}{\rm (2 \ jet)}$. In this paper, we adopt minimal 
  cuts to the jet rapidity $\eta$ and transverse momentum $p_T$ so that $\vert \eta_i \vert <2.5$,  $p_{T1}>100$GeV, and $p_{T2}>50$GeV. 

The minimization is taken over all $p_1$ and $p_2$  that satisfy $p^T_{1}+p^T_{2}=p^T_{miss}$  and $(p_1)^2=(p_2)^2=m^2_{\rm test}$.

We can define $p_{\rm rec}^{(1)}$ and $p_{\rm rec}^{(2)}$ from the $p^{T}_{1}({\rm min})$ and  $p^{T}_{2}({\rm min})$ which minimize the above equation. The conditions to solve $p_{\rm rec}^{(i)}$ are (In our case, $Q_1$ and $Q_2$ are the intermediate massive heavy quark partners)
\begin{eqnarray}
P_{Q_1}^2& \equiv &(p^{(1)}_{\rm vis}+p_{\rm rec}^{(1)})^2=m_Q^2, \cr
P_{Q_2}^2&\equiv &(p^{(2)}_{\rm vis}+p_{\rm rec}^{ (2)})^2
=m_Q^2, \cr
(p^{(1)}_{\rm rec})^2&=&(p^{(2)}_{\rm rec})^2=m^2_{\rm test},
\end{eqnarray}
for the transverse momentum 
\begin{equation}
p^{(1)T}_{\rm rec}\equiv  p^{T}_1({\rm min}) ,\ \ p^{(2)T}_{\rm rec}\equiv p^{T}_2(\rm min).
\end{equation}
For the events with $m_{T2}\sim m^{\rm max}_{T2}$, $p^{T}_1(\rm min)$ and
 $p^{T}_2({\rm min})$ are very close to the true $B_H$ transverse  momenta. Therefore 
 we define the reconstructed azimuthal angle difference $\phi({\rm rec})$ 
 based on $P_{Q_1}$ and  $P_{Q_2}$, namely 
 $\phi_{\rm rec}$ is the  azimuthal angle difference in the CM frame of $P_{Q_1}+P_{Q_2}$, where the  $z$-axis is along the direction of the momenta $P_{Q}$. There are two undetermined components in the $p^{(i)}_{rec}$ and we have the two mass-shell equations for both sides of the 
  decays, which give $2\times 2=4$ solutions for $(p^{(1)}_{\rm rec}, p^{(2)}_{\rm rec})$. 
  Two of them give trivial solutions with $\phi({\rm rec})=0$ , and we study 
the other two non-trivial solutions, which lead to two azimuthal 
angle solutions, $\phi({\rm rec},1)$ and $\phi({\rm rec},2)$.

\begin{figure}
\begin{center}
\includegraphics[width=7cm]{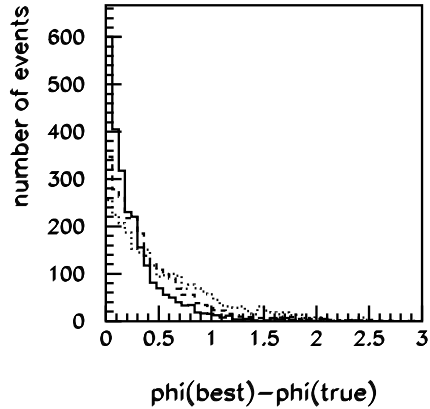}
\caption{The difference between reconstructed azimuthal angle difference 
and  $\phi(\rm true)$ at parton level. The solid line shows the difference from 
the best solution, the thick dashed histogram plots  both of the reconstructed solutions, 
and the thin dotted line is the difference to the lab frame azimuthal angle.  All histograms corresponds 
to 20000 events before the cuts. }\label{compare}
\end{center}
\end{figure}

 The merit of using MAOS reconstruction is limited. In Fig.~\ref{compare} 
 we plot the difference between the reconstructed $\phi({\rm rec},i)$ of the parton level event  and the true parton level azimuthal angle difference $\phi({\rm true}) $ for the events near the end point with $m_{T2}> 500$~GeV. The plot is based on 20000 generated events where $\sim$ 2500 passed through the cut. 
  Here the  solid histogram is  $\Delta\phi({\rm best})=\min_{i=1,2}
\vert (\phi({\rm rec},i)-\phi({\rm true}))\vert $, 
  the  dark dotted histogram contains  both of $\Delta\phi(i)\equiv \vert\phi({\rm rec},i)-\phi({\rm true) } \vert$ 
  (scaled by a factor of 0.5), and 
 the  thin dotted histogram is  $\Delta\phi({\rm Lab})=\vert \vert\phi_1-\phi_2\vert-\phi (\rm true) \vert$, where $\phi_i$ is the azimuthal angle in the Lab frame. The matching between $\phi(\rm true)$ and  $\phi(\rm best)$ is good. However, once we take into account both of the solutions, $\phi({\rm rec},1)$ and $\phi({\rm rec},2)$, the improvement from the simple Lab frame quantity $\Delta\phi({\rm Lab} ) $ is moderate. The mean value of $\Delta\phi(\rm best)$ is 0.28, while  it is 0.41 for $\Delta\phi({\rm rec},  i)$ and 0.50  for $\Delta\phi(\rm Lab)$ respectively.  
 
\begin{figure} 
\begin{center}
\includegraphics[width=4.5cm]{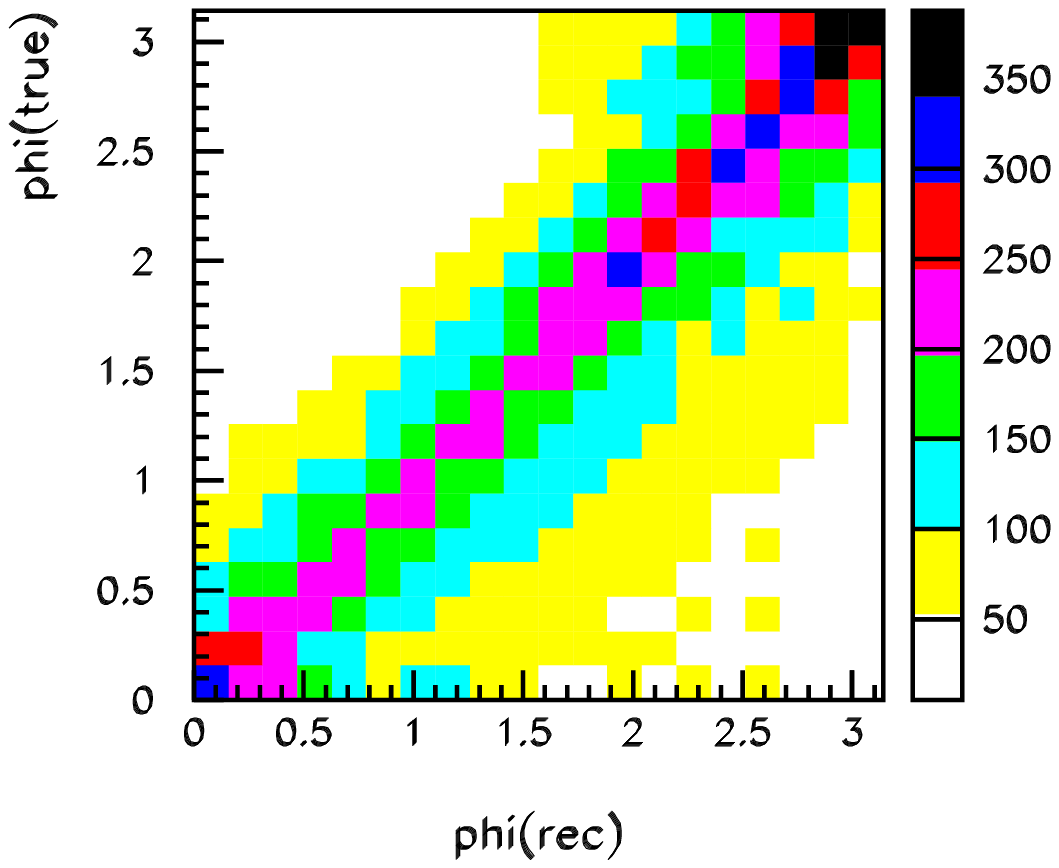}
\includegraphics[width=4.5cm]{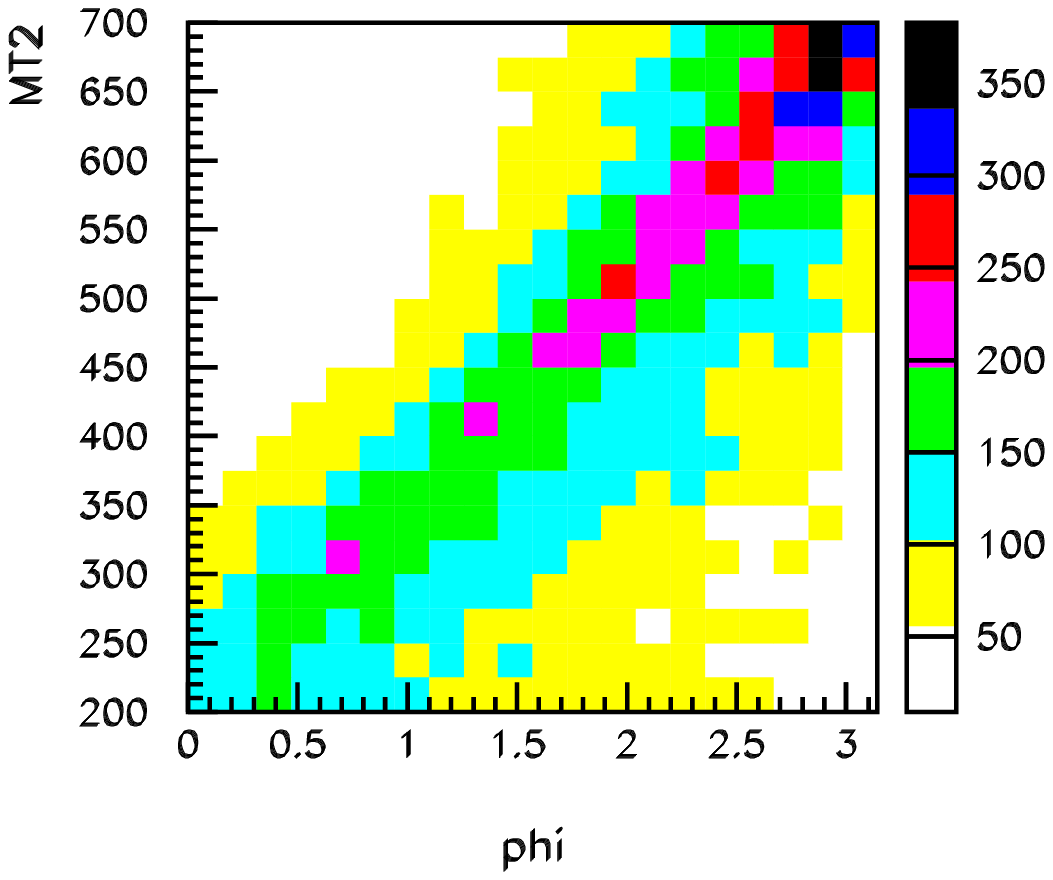}
\includegraphics[width=4.5cm]{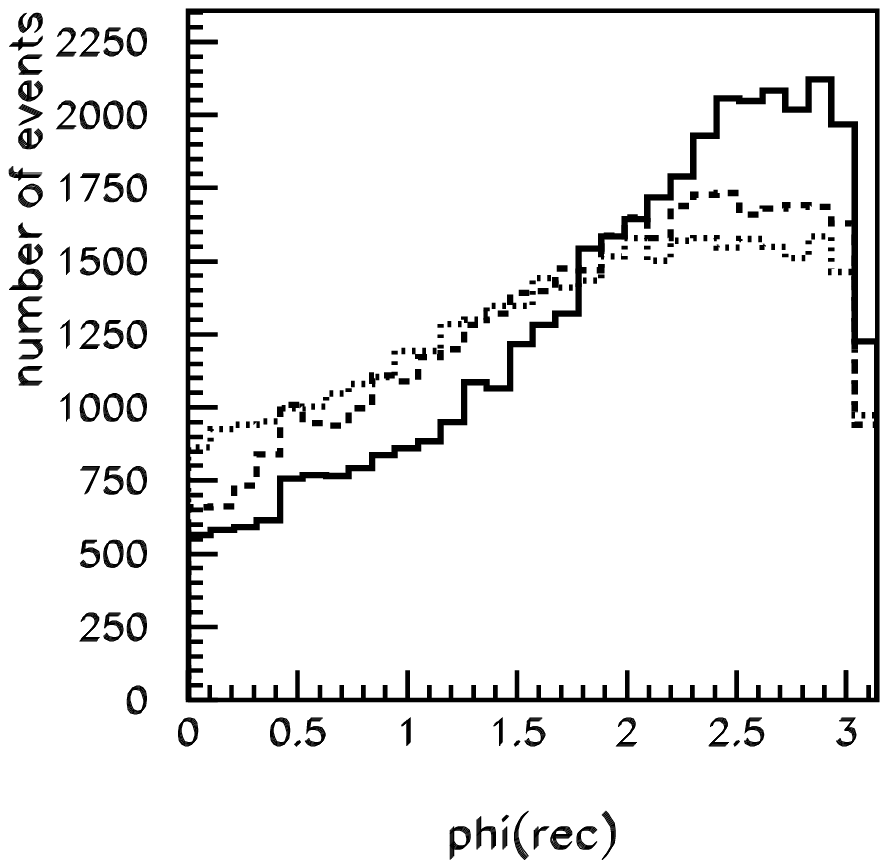}
\end{center}
\caption{Left: parton level $\phi(\rm rec)$ vs. $\phi(\rm true)$ distribution at Point A. Middle;  jet level $\phi(\rm rec)$ vs. $\phi(\rm true)$ distribution at Point A.  Right $\phi(\rm rec)$ distributions at jet level (solid histogram), jet level without spin correlation (thick dashed histogram), parton level without spin correlation (thin dashed histogram). See the 
explanation in the text for details. Each plot corresponds to $2\times 10^4$ events before the cuts.  
 } \label{fig7}
\end{figure}

At jet level, we have to consider the possibility that the selected momenta do not match with parton level momenta.  In Fig.~\ref{fig7}(left)  and Fig.~\ref{fig7}(middle),  we compare parton level and jet level distributions in $\phi(\rm rec)$-$\phi(\rm true)$ plain. 
In the left plot, we plot $\phi({\rm rec})$ with $\phi(\rm true)$ at parton level, and in the middle plot $\phi{(\rm rec)}$  is calculated for the two highest $p_T$ jets in the events  after the detector smearing. For $\phi (\rm true)\sim 0$ the probability to mis-reconstruct the azimuthal angle is 
 high. This is because when two partons go collinear, the correlation between parton momenta and  jet momenta is not good.  In Fig~\ref{fig7}, we compare the distribution of $\phi ({\rm  rec})$ at Point A (solid line) with the distribution of the events without spin correlation (a thick dashed histogram) and 
 the parton level distribution without spin correlation (a thin dashed histogram). 
All distributions have a peak at $\phi=\pi$, but the distribution with 
spin correlation is most strongly peaked at $\phi ({\rm  rec}) \sim \pi$ as expected.

\begin{figure} 
\begin{center}
\includegraphics[width=4.5cm]{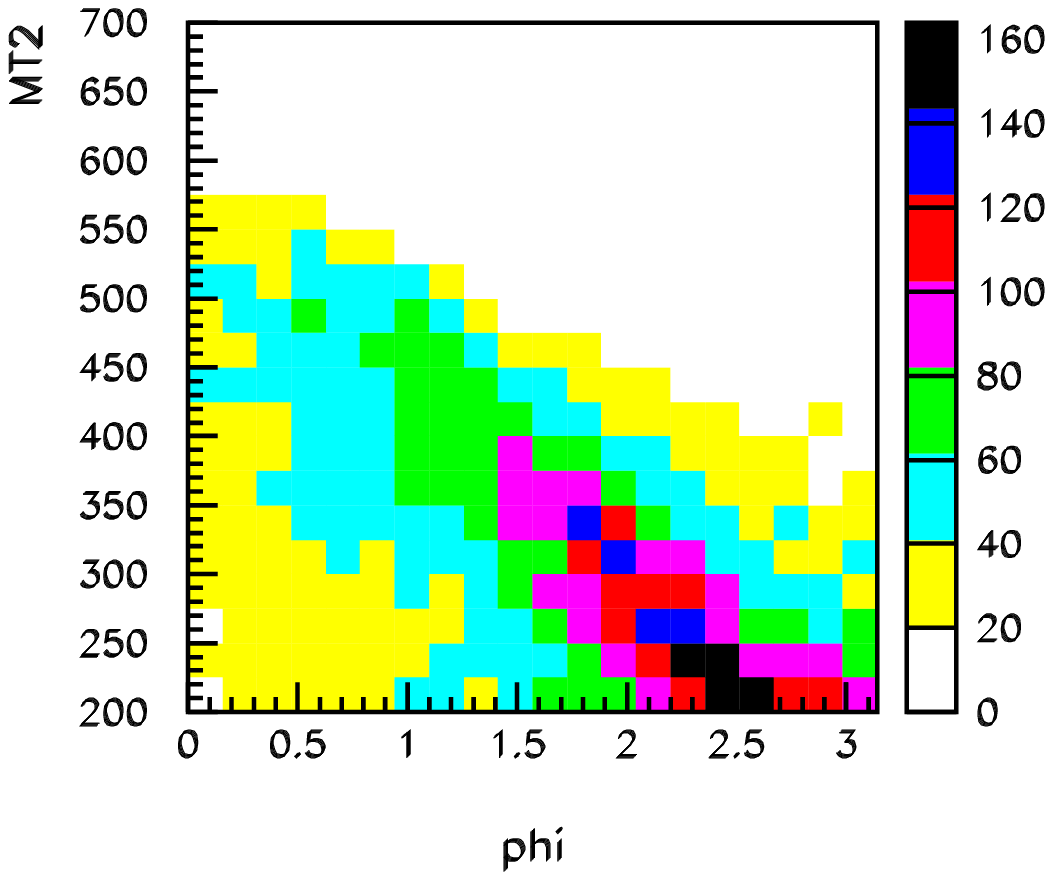}
\includegraphics[width=4.5cm]{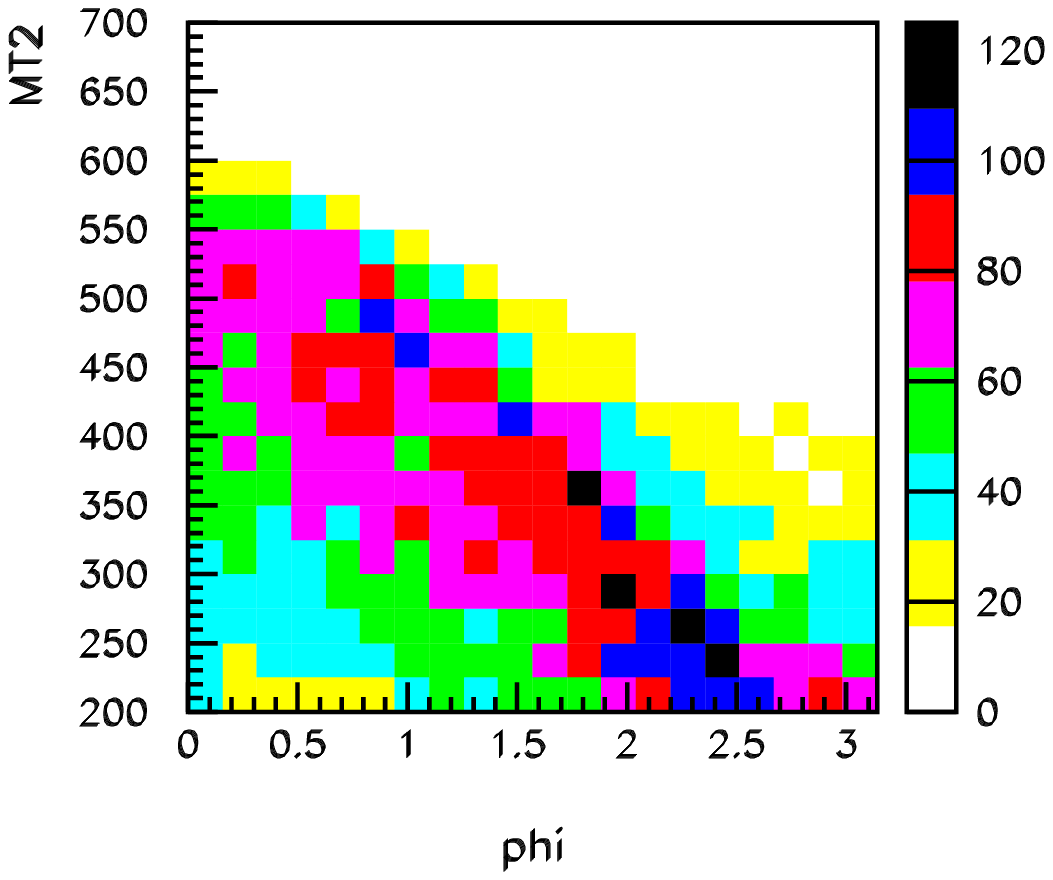}
\includegraphics[width=4.5cm]{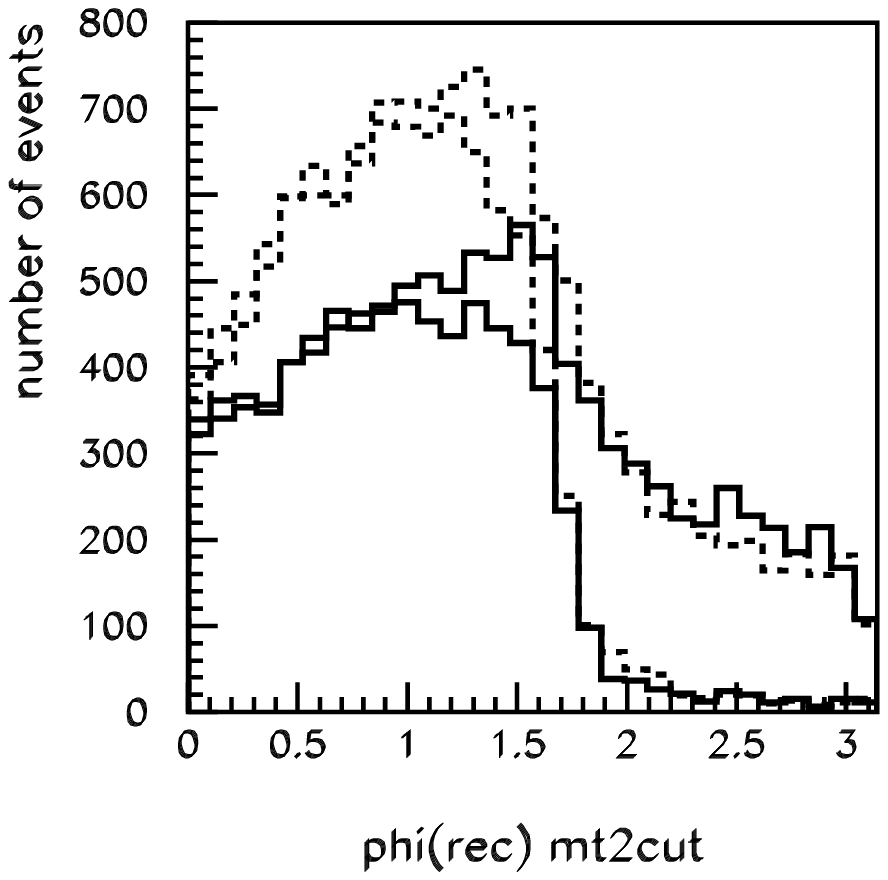}
\end{center}
\caption{The $\phi({\rm {rec}})$ vs $m_{T2}$ distributions with and without spin correlation (left and middle plots, respectively) Right: The $\phi({\rm {rec}})$ distribution with spin correlation (solid histogram) and without spin correlation (dashed histogram). Each plots generates $2\times 10^4$ events before the cuts. } 
\label{fig8}
\end{figure}

The distribution of $\phi(\rm rec)$ is significantly altered by the kinematical cuts to reduce the background. To explain the effect of the cuts, we show the correlation between $\phi$(true) and $m_{T2}$ in Fig.~ \ref{fig8}. In Fig. \ref{fig8} left (middle)  we show the distribution with (without) spin correlation. Here, $m_{T2}$ is calculated for the two highest $p_T$ jets as the visible objects of the events, and $m_\textrm{test}=m_{B_H}$. We plot both of the reconstructed azimuthal angle solutions in the plot. The expected endpoint of the $m_{T2}$ distribution is 600~GeV.  For the two jets + missing $E_T$ signature, it is known that requiring a lower limit on the $m_{T2}$ value is an efficient cut to remove the standard model background \cite{Barr:2009wu}. In the region $\phi\sim 0 (\pi)$, $m_{T2}$ tends to be large (small). At Point A the number of events at $\phi_{rec}\sim 0$ is small compared with the distribution without spin correlation. 

In Fig. \ref{fig8} (right), we show the 
$\phi({\rm {rec}})$ distribution with various $m_{T2}$ cuts. 
Solid and dotted lines with a sharp cutoff around $\phi(\rm{rec})<2$ 
correspond to $m_{T2}({\rm 2\ jet} ) >400$~GeV, where $m_{T2}({\rm 2\ jet})$ is the $m_{T2}$ calculated with the two highest $p_T$ jets as the visible objects. 
About 3600 events remain after the cut for spin correlated events and 5000 events 
for spin non-correlated events. 
Significant suppression  is observed for 
$\phi({\rm rec}) \lesssim 2$ for the events with spin correlation. The sharp edge at $\phi 
\sim 2$ appears because the events above $\phi(\rm true) \gtrsim 2$ for $m_{T2}>400$~GeV are kinematically forbidden.

The other two histograms with significant tails beyond $\phi(\rm{rec}) \gtrsim 2$ are for the events with $m_{T2}>$400~GeV, but the $m_{T2}$ is calculated for all jets in the events using inclusive $m_{T2}$ defined in \cite{Nojiri:2008hy, Nojiri:2008vq}. Here the jets and leptons in an event are clustered into 
two visible objects to obtain $m_{T2}$. The number of events remaining after the cut is $\sim$ 5300 for spin correlated events and 
$\sim$ 6500 for spin non-correlated events. 
The tail events are the $QQ$ events  with significant initial state radiation.  
The events with $m_{T2}({\rm 2\  jet}) <400$~GeV contaminate the $\phi(\rm{rec}) > 2$ region due to the initial state radiation. 

The number of events 
with $\phi(\rm{rec})>2$ is roughly the same between 
spin correlated and non-correlated samples, because  
ISR and  the leading jets from $Q$ decay are not spin correlated. 
This  is encouraging because
the events in this region may be regarded as control samples 
to determine the total number of produced events. 
Together with the number of events with $\phi(\rm{rec}) < 2$, 
we may be able to determine how much azimuthal angle correlation 
exists in the events. To judge the usefulness of the method, it is necessary to conduct 
a more careful study of the dependence on more general production and decay patterns, and the  effects of other cuts such as missing 
$E_T$. 

\section{Discussion and conclusion}
\label{sec:conclusion}

In the coming years, the LHC will accumulate more events, and the signature 
of supersymmetry, namely an excess of events with $\sla{E}_T$, may be found. However, there are other models with similar signatures but different spin structures. In this paper 
we discuss a way to measure the spin dependence of SUSY-like signatures, 
especially those that might be observed in the early stages of the ATLAS and CMS experiments. 

We focus on the spin correlation in the 2 high $p_T$ jets + missing $E_T$ $+ X$  signature at the LHC. The signature would arise from a $ p p \rightarrow QQ$ process when 1) the model has a colored quark partner $Q$ that directly couples to quarks. 2) The colored particle dominantly decays into a lighter parity odd particle $\chi$ and a quark $q$. The $\chi$ are gauginos or Higgsinos in SUSY or heavy gauge partners in same spin partner (LHT or UED like) models.  3) The mass  satisfies 
$m_Q\gg m_{\chi}$.  The $\chi$ may  further decay into other particles involving 
the lightest ``parity odd" particles. In that case, the jets from $Q$ decays are prominent 
among the jets and initial state radiation \cite{Nojiri:2010mk}. Depending on the spin of $Q$ and its polarization, the two jets show non-trivial spin correlations, both in the polar angle dependence of  $Q\rightarrow q \chi$ decay and in the azimuthal angle correlation of the two quarks. The effect can be seen in $E_{\rm Tmiss}$ vs $M_{\rm eff}$ distributions, and the reconstructed azimuthal angle correlation using MAOS momentum. 

The expected spin correlation depends strongly on the production process. 
In this paper we have studied the production process based on the $t$ channel 
exchange of the heavy gauge bosons expected in same spin partner (LHT or UED like) models. In those models a parity odd quark $Q$ couples directly to a quark $q$, and the interaction among $Q$, $q$ and parity odd heavy gauge bosons $B_H$ and $G_H$ could be chiral. We have shown that the helicity structure of the amplitude is very sensitive to the mass and the interaction of the heavy gauge bosons. The enhanced helicity $h=0$ component plays an especially important role. As an example, we show the distributions of $uu\rightarrow U^{(R)} U^{(R)}$. The polar angle dependence comes from overall polarization, while  azimuthal angle correlation comes from  interference between spin flipping $B_H$ exchange and spin conserving heavier $G_H$ exchange. The azimuthal angle correlation is visible because the amplitude is forwardly peaked due to the  $B_H$ exchange contribution. 

Due to the azimuthal angle correlation of the events, the number of events 
may be suppressed near the $m_{T2}$ (2 jet) end points. The high $m_{T2}$ region is the 
signal region because the SM background is smaller, therefore spin 
correlations affect the estimation of the total production cross section. 
 We note that there are phase 
space boundaries of the events of 
 $pp\rightarrow QQ\rightarrow uu+ E_{\rm Tmiss}$. 
Some events can lie outside 
the phase space due to ISR, and they are  less affected 
by the spin correlations.  We note that such events may be actively used 
to estimate the total cross section of the events. 

In this paper, we do not study the corresponding distribution in SUSY models in detail. In supersymmetric models we do not expect spin correlations because the partner of the  quark has spin 0. This leads to a high $E_{\rm Tmiss}/M_{\rm eff}$, a hard $m_{T2}$ end point, and a flatter reconstructed azimuthal angle correlation for two jets. Of course there are other important differences.  the squark squark pair production is generally small in SUSY-like models, and gluino squark co-production is dominant. The gluinos may decay into squark and quark, and 
the ISR of the process involving the gluino is larger, so selecting the correct jets coming from squark decay may not be straightforward. However, it is shown in Ref. \cite{Nojiri:2010mk} that the two highest $p_T$ jets in the events are likely from $ \tilde{q} \rightarrow \chi q $ decay. The inclusive  $m_{T2}$ distributions with high $p_T$ jets show a sharp end point for the model parameter $m_{\tilde{q}}<m_{\tilde{g}}$. This means the two jets 
from squark decays are the dominant part of the event activity. Therefore we believe the feature discussed in this paper would be useful for SUSY studies as well.
 
For extracting the physics behind the signatures at the LHC, using fully spin correlated amplitudes 
for production and decay is essential. To reproduce the structure of the amplitude, the amplitude of $qq \rightarrow QQ \rightarrow qq B_H B_H + X$ should be fully calculated. In this paper we calculated the full amplitude using Madgraph \cite{Alwall:2007st}, but other amplitude calculators or generators such as Comphep/Calchep \cite{calchep} and Herwig++ \cite{Bahr:2008pv} should be able to reproduce the effect. On the other hand, when we interface the two to two process to PYTHIA \cite{Sjostrand:2006za}, the amplitude does not have the proper spin correlation, and therefore it is not appropriate for this study. 

\section*{Acknowledgments}
We would like to thank Won Sang Cho, William Klemm, Matthew Sudano and Yuji Tachikawa for useful discussions. This work was supported by the World Premier International Research Center Initiative (WPI initiative) by MEXT, Japan. The work of M. N. and J. S. was also supported by the Grant-in-Aid for scientific research (22540300 for M.N. and Young Scientists (B) 21740169 for J.S. ) from Japan Society for Promotion of Science. M. N. would also like to thank SLAC for their hospitality where part of the research was conducted. The model file of Madgraph will be provided upon request. (Mihoko M. Nojiri, nojiri@post.kek.jp).
	


\newpage
\begin{thebibliography}{99}	
\bibitem{Cheng:2004yc}
  H.~C.~Cheng and I.~Low,
  JHEP {\bf 0408}, 061 (2004)
  [arXiv:hep-ph/0405243].

\bibitem{Appelquist:2000nn}
  T.~Appelquist, H.~C.~Cheng and B.~A.~Dobrescu,
  Phys.\ Rev.\  D {\bf 64}, 035002 (2001)
  [arXiv:hep-ph/0012100].

\bibitem{Cheng:2002ab}
  H.~C.~Cheng, K.~T.~Matchev and M.~Schmaltz,
  Phys.\ Rev.\  D {\bf 66}, 056006 (2002)
  [arXiv:hep-ph/0205314].
  
\bibitem{Cheng:2002iz}
  H.~C.~Cheng, K.~T.~Matchev and M.~Schmaltz,
  Phys.\ Rev.\  D {\bf 66}, 036005 (2002)
  [arXiv:hep-ph/0204342].

\bibitem{Hubisz:2008gg}
 J.~Hubisz, J.~Lykken, M.~Pierini and M.~Spiropulu,
 Phys.\ Rev.\  D {\bf 78} (2008) 075008
 [arXiv:0805.2398 [hep-ph]].

\bibitem{Hallenbeck:2008hf}
  G.~Hallenbeck, M.~Perelstein, C.~Spethmann, J.~Thom and J.~Vaughan,
  Phys.\ Rev.\  D {\bf 79}, 075024 (2009)
  [arXiv:0812.3135 [hep-ph]].

\bibitem{Kane:2008kw}
  G.~L.~Kane, A.~A.~Petrov, J.~Shao and L.~T.~Wang,
  J.\ Phys.\ G {\bf 37}, 045004 (2010)
  [arXiv:0805.1397 [hep-ph]].
    
\bibitem{Barr:2004ze}
  A.~J.~Barr,
  Phys.\ Lett.\  B {\bf 596} (2004) 205
  [arXiv:hep-ph/0405052].
\bibitem{Goto:2004cpa}
  T.~Goto, K.~Kawagoe and M.~M.~Nojiri,
  Phys.\ Rev.\  D {\bf 70} (2004) 075016
  [Erratum-ibid.\  D {\bf 71} (2005) 059902]
  [arXiv:hep-ph/0406317].

\bibitem{Smillie:2005ar}
  J.~M.~Smillie and B.~R.~Webber,
  JHEP {\bf 0510}, 069 (2005)
  [arXiv:hep-ph/0507170].

\bibitem{Datta:2005zs}
  A.~Datta, K.~Kong and K.~T.~Matchev,
  Phys.\ Rev.\  D {\bf 72} (2005) 096006
  [Erratum-ibid.\  D {\bf 72} (2005) 119901]
  [arXiv:hep-ph/0509246].
\bibitem{Athanasiou:2006ef}
  C.~Athanasiou, C.~G.~Lester, J.~M.~Smillie and B.~R.~Webber,
  JHEP {\bf 0608} (2006) 055
  [arXiv:hep-ph/0605286].

\bibitem{Wang:2006hk}
  L.~T.~Wang and I.~Yavin,
  JHEP {\bf 0704} (2007) 032
  [arXiv:hep-ph/0605296].
  
\bibitem{Burns:2008cp}
 M.~Burns, K.~Kong, K.~T.~Matchev and M.~Park,
 JHEP {\bf 0810} (2008) 081
 [arXiv:0808.2472 [hep-ph]].

\bibitem{Buckley:2007th}
  M.~R.~Buckley, H.~Murayama, W.~Klemm and V.~Rentala,
  Phys.\ Rev.\  D {\bf 78} (2008) 014028
  [arXiv:0711.0364 [hep-ph]].

\bibitem{Buckley:2008pp}
  M.~R.~Buckley, B.~Heinemann, W.~Klemm and H.~Murayama,
  Phys.\ Rev.\  D {\bf 77} (2008) 113017
  [arXiv:0804.0476 [hep-ph]].

\bibitem{Boudjema:2009fz}
 F.~Boudjema and R.~K.~Singh,
 JHEP {\bf 0907} (2009) 028
 [arXiv:0903.4705 [hep-ph]].

\bibitem{Park:2009cs}
  S.~C.~Park and J.~Shu,
  Phys.\ Rev.\  D {\bf 79}, 091702 (2009)
  [arXiv:0901.0720 [hep-ph]].
  
\bibitem{Flacke:2008ne}
  T.~Flacke, A.~Menon and D.~J.~Phalen,
  Phys.\ Rev.\  D {\bf 79}, 056009 (2009)
  [arXiv:0811.1598 [hep-ph]].
      
\bibitem{Agashe:2007jb}
  K.~Agashe, A.~Falkowski, I.~Low and G.~Servant,
  JHEP {\bf 0804}, 027 (2008)
  [arXiv:0712.2455 [hep-ph]].
     
 \bibitem{ArkaniHamed:2001ca}
  N.~Arkani-Hamed, A.~G.~Cohen and H.~Georgi,
  Phys.\ Rev.\ Lett.\  {\bf 86}, 4757 (2001)
  [arXiv:hep-th/0104005]; 
\bibitem{Hill:2000mu}
  C.~T.~Hill, S.~Pokorski and J.~Wang,
  Phys.\ Rev.\  D {\bf 64}, 105005 (2001)
  [arXiv:hep-th/0104035].

\bibitem{'tHooft:1972fi}
  G.~'t Hooft and M.~J.~G.~Veltman,
  Nucl.\ Phys.\  B {\bf 44}, 189 (1972).

\bibitem{Cornwall:1974km}
  J.~M.~Cornwall, D.~N.~Levin and G.~Tiktopoulos,
  Phys.\ Rev.\  D {\bf 10}, 1145 (1974)
  [Erratum-ibid.\  D {\bf 11}, 972 (1975)].

\bibitem{Alwall:2007st}
  J.~Alwall {\it et al.},
  JHEP {\bf 0709} (2007) 028
  [arXiv:0706.2334 [hep-ph]].

\bibitem{Meade:2007js}
  P.~Meade and M.~Reece,
  arXiv:hep-ph/0703031.

\bibitem{Sjostrand:2006za}
  T.~Sjostrand, S.~Mrenna and P.~Z.~Skands,
  JHEP {\bf 0605} (2006) 026
  [arXiv:hep-ph/0603175].

\bibitem{RichterWas:2002ch}
  E.~Richter-Was,
  arXiv:hep-ph/0207355.

\bibitem{Salam:2007xv}
  G.~P.~Salam and G.~Soyez,
  JHEP {\bf 0705} (2007) 086
  [arXiv:0704.0292 [hep-ph]].
 See also, http://www.lpthe.jussieu.fr/~salam/fastjet/


\bibitem{Cho:2008tj}
  W.~S.~Cho, K.~Choi, Y.~G.~Kim and C.~B.~Park,
  Phys.\ Rev.\  D {\bf 79} (2009) 031701
  [arXiv:0810.4853 [hep-ph]].

\bibitem{Cho:2009wh}
  W.~S.~Cho, K.~Choi, Y.~G.~Kim and C.~B.~Park,
  Nucl.\ Phys.\ Proc.\ Suppl.\  {\bf 200-202} (2010) 103
  [arXiv:0909.4853 [hep-ph]].
  
\bibitem{Barr:2003rg}
  A.~Barr, C.~Lester and P.~Stephens,
  J.\ Phys.\ G {\bf 29} (2003) 2343
  [arXiv:hep-ph/0304226].
  
\bibitem{Barr:2009wu}
  A.~J.~Barr and C.~Gwenlan,
  Phys.\ Rev.\  D {\bf 80} (2009) 074007
  [arXiv:0907.2713 [hep-ph]].

 \bibitem{Nojiri:2008hy}
  M.~M.~Nojiri, Y.~Shimizu, S.~Okada and K.~Kawagoe,
  JHEP {\bf 0806} (2008) 035
  [arXiv:0802.2412 [hep-ph]].

\bibitem{Nojiri:2008vq}
  M.~M.~Nojiri, K.~Sakurai, Y.~Shimizu and M.~Takeuchi,
  JHEP {\bf 0810} (2008) 100
  [arXiv:0808.1094 [hep-ph]].

\bibitem{Nojiri:2010mk}
  M.~M.~Nojiri and K.~Sakurai,
  arXiv:1008.1813 [hep-ph]
  
\bibitem{calchep}
 A.~Pukhov,
  arXiv:hep-ph/0412191.
\bibitem{Bahr:2008pv}
  M.~Bahr {\it et al.},
  Eur.\ Phys.\ J.\  C {\bf 58} (2008) 639
  [arXiv:0803.0883 [hep-ph]].


\end{thebibliography}
\end{document}